\documentclass[]{SolarPhysics}
\usepackage[optionalrh,solaenum]{spr-sola-addons} 
\usepackage{solaheader}
\newcommand{\solareceiveddate}{14 July 2010} 
\newcommand{\solaaccepteddate}{29 December 2010}

\usepackage[usenames]{color}
\usepackage{graphicx}
\usepackage{bm}
\usepackage{url}                         

\def\rsun{$R_{\odot}$}
\def\rss{$R_{\rm SS}$}
\def\mrSun{R_{\odot}} 
\def\deg{$^\circ$}
\def\mdeg{^\circ}

\def\bb{b}
\def\momNe{\left< N_e^2\right>^{1/2}}
\def\Ne{N_\mathrm{e}}
\def\WT{W_T}

\def\Tm{T_\mathrm{m}}
\def\Tions{T_{\rm ions}}
\def\aTm{\left<\Tm\right>}

\def\etal{\textit{et al.}}

\def\Tfit{T_{\rm fit}}
\def\Te{T_{\rm e}}
\def\TH{T_{\rm H}}
\def\THe{T_{\rm He}}
\def\l{\lambda_{{\rm N}}}
\def\s2Tm{\sigma^2_{\Tm}}
\def\ssWT{\sigma^2_{\WT}}
\def\dT{\Delta T}
\def\RR{R^2}
\def\gSun{g_{\rm s}}
\def\MSun{M_{\rm s}}
\def\R#1{$R_{#1}$}

\begin{document}
\begin{article}
\begin{opening}

\title{The WHI Corona from Differential Emission Measure Tomography}
\runningtitle{WHI DEM Tomography}
\runningauthor{A.M.~V\'asquez \textit{et al.} }

\author{Alberto~M.~\surname{V\'asquez}$^{1}$\sep
Zhenguang~\surname{Huang}$^{2}$ \sep 
Ward~B.~\surname{Manchester IV}$^{2}$ \sep
Richard~A.~\surname{Frazin}$^{2}$
}

\institute{
$^{1}$
Instituto de Astronom\'{\i}a y F\'{\i}sica del Espacio
(CONICET-UBA) and FCEN (UBA), CC 67 - Suc 28, Ciudad de Buenos Aires, Argentina;
email: \url{albert@iafe.uba.ar}\\
$^{2}$
Dept. of Atmospheric, Oceanic and Space Sciences, University of
Michigan, Ann Arbor, MI 48109; 
email: \url{zghuang@umich.edu, chipm@umich.edu, rfrazin@umich.edu}
}

\begin{abstract}
A three dimensional (3D) tomographic reconstruction of the local differential
emission measure (LDEM) of the global solar corona during the whole heliosphere
interval (WHI, Carrington rotation CR-2068) is presented, based on STEREO/EUVI
images. We determine the 3D distribution of the electron density, mean
temperature, and temperature spread, in the range of heliocentric heights 1.03
to 1.23\,\rsun. The reconstruction is complemented with a potential field source
surface (PFSS) magnetic-field model. 
The streamer core, streamer legs, and subpolar regions are analyzed and compared
to a similar analysis previously performed for CR-2077, very near the absolute
minimum of the Solar Cycle 23. In each region, the typical values of density and
temperature are similar in both periods. The WHI corona exhibits a streamer
structure of relatively smaller volume and latitudinal extension than during
CR-2077, with a global closed-to-open density contrast about 6\% lower, and a
somewhat more complex morphology. 
The average basal electron density is found to be about $2.23$ and $1.08\times
10^8\,{\rm cm^{-3}}$, in the streamer core and subpolar regions, respectively.
The electron temperature is quite uniform over the analyzed height range, with
average values of about 1.13 and 0.93 MK, in the streamer core and subpolar
regions, respectively. {Within the streamer closed region, both periods show
higher temperatures at mid-latitudes and lower temperatures near the equator.}
Both periods show $\beta>1$ in the streamer core and $\beta<1$ in the
surrounding open regions, with CR-2077 exhibiting a stronger contrast.
Hydrostatic fits to the electron density are performed, and the scale height is
compared to the LDEM mean electron temperature. Within the streamer core, the
results are consistent with an isothermal hydrostatic plasma regime, with the
temperatures of ions and electrons differing by up to about 10\%. In the
subpolar open regions, the results are consistent with departures from thermal
equilibrium with $T_{\rm ions}>\Te$ (and values of $T_{\rm ions}/\Te$ up to
about 1.5), and/or the presence of wave pressure mechanisms linear in the
density. 
\end{abstract}

\keywords{Corona, Quiet; Magnetic fields, Corona; Tomography; Differential
Emission Measure; EUV Imaging; STEREO Mission}
\end{opening}

\section{Introduction}

Advancing our understanding of the processes that heat and accelerate the
coronal plasma now requires empirical knowledge of its three-dimensional (3D)
structure.  Coronal images  are two-dimensional projections of the 3D structure,
and a number of methods have been used to recover the 3D information from the 2D
images. Available techniques include a variety of approaches, with diverse aims,
strengths, and limitations. Towards this end, solar rotational tomography (SRT)
constitutes a powerful empirical technique. Since the original work by
Altschuler and Perry (1972), SRT has been developed and applied to polarized
white-light image time series, allowing for the reconstruction of the 3D
structure of the coronal electron number density. A modern implementation of SRT
can be found in Frazin (2000) and Frazin and Janzen (2002), and a comprehensive
review of its development in Frazin and Kamalabadi (2005).

One of the primary goals of NASA's dual-spacecraft \textit{Solar Terrestrial
Relations Observatory} (STEREO) mission is precisely to determine the 3D
structure of the corona (Kaiser \emph{et al.}, 2008). The \textit{Extreme
UltraViolet Imager} (EUVI) on the STEREO mission returns high-resolution
($1.6''$) narrow-band images centered over Fe emission lines at 171, 195, 284
\AA, and the He\,{\sc ii} 304 \AA\ line (Howard \emph{et al.}, 2008). In this
context, we have developed a novel technique, named differential emission
measure tomography (DEMT). The technique was theoretically proposed by Frazin
\etal{} (2005), and fully developed and applied to STEREO/EUVI data by Frazin
\etal{} (2009; henceforth FVK09). DEMT takes advantage of the solar rotation to
provide the multiple views required for tomography, as well as of the dual view
angles provided by the STEREO spacecraft, the use of which allows for a reduced
data-gathering time. Based on the input of EUV-image time series, DEMT produces
maps of the 3D EUV emissivity, and a 3D DEM analysis free of 2D projection
effects. As explained in FVK09, the first three moments of this local DEM (or
LDEM) analysis give 3D maps of the electron density, the mean electron
temperature, and the electron temperature spread. A major advantage of DEMT is
that it obviates the need for \textit{ad-hoc} modeling of specific structures of
interest. Its main (current) limitation is the assumption of a static corona
during the data-gathering process, implying that the reconstructions are
reliable only in coronal regions populated by structures that are stable
throughout their disk transit in the images.  In contrast to other approaches,
DEMT does not require background subtraction, and is global (\textit{i.e.} it
considers the entire corona), but it does not resolve individual loops.

In V\'asquez \etal{} (2009), we published the first empirically derived 3D
density and temperature structure of coronal-filament cavities, structures that
are particularly interesting to study as filament eruptions are the progenitors
of about 2/3 of all CMEs (Gibson \etal, 2006). In V\'asquez \etal{} (2010, VFM10
hereafter), we presented the first EUVI/STEREO DEMT analysis of the global
corona, specifically for the period CR-2077, belonging to the Solar Cycle 23
extended solar-activity minimum period. In the present work, we develop a
similar DEMT analysis for EUVI/STEREO data corresponding to the Whole
Heliosphere Interval (WHI) period CR-2068 (20 March 2008, 01:14 UT through 16
April 08:05 UT). We also show a potential field source surface (PFSS)
magnetic-field model based on the \textit{Michelson Doppler Imager} (MDI/SOHO)
magnetograms of the same period.

The \emph{Comparative Solar Minima} (CSM) working group (WG), sponsored by
Division II (Sun and Heliosphere) of the International Astronomical Union (IAU),
focuses on the research of the coupled Sun--Earth system during solar minimum
periods. It seeks to characterize the system at its most basic, ``ground state"
and aims to understand the degree and nature of variations within and between
solar minima. In this context, we present here the results of the DEMT+PFSSM
analysis of the WHI period. We discuss the implication of our model for the
thermodynamical structure of the equatorial streamer belt, as well as for the
surrounding magnetically open regions, both at the latitudes of the so-called
``streamer legs'', and at the higher subpolar latitudes. To address the
central interests of the IAU/CSM WG, we compare our results both with our
similar
analysis of CR-2077 (VFM10), as well as with results from studies of the Whole
Sun Month period (WSM, CR-1913, 22 August  through 18 September 1996), belonging
to the previous solar-cycle minimum.

\section{Summary of Differential Emission Measure Tomography}\label{DEMTsec}

We summarize in this section the main aspects of the DEMT methodology, a
comprehensive description of the technique can be found in FVK09 and VFM10. DEMT
consists of two phases. A first one applies solar rotational tomography (SRT) to
a series of EUV images, the $K$ instrumental bands independently. As a result, 
the values of the $K$ \emph{filter band emissivities} (FBEs) $\zeta_{k,i}$ are
obtained at each tomographic grid cell (or voxel) $i$. The solution of the
problem involves the application of \emph{regularization} (or smoothing) methods
to stabilize the inversion. The strength of the regularization is controlled by
a single parameter [$p$], which is determined via the statistical procedure of
\emph{cross validation}.

In the second phase, a local DEM analysis is performed at each voxel using the
local FBE values and assuming an optically thin plasma emission model, such as
CHIANTI, for the computation of the different bands instrumental temperature
responses. As a result, the LDEM distribution [$\xi_i(T)$] is obtained at each
voxel [$i$]. The LDEM zeroth through second moments (Equations (4) through (6)
in
VFM10) give the voxels' mean squared electron density [$N_{e,i}^2$], mean
electron temperature [$T_{m,i}$], and squared electron temperature spread
[$W_{T,i}^2$], respectively.

As the EUVI coronal bands are dominated by iron lines, their temperature
responses are proportional to that element's abundance. The root-mean-squared
electron density [$N_{e,i}$] derived at each voxel is then inversely
proportional to the squared root of the Fe abundance, assumed in this work to be
uniform and equal to $\mathrm{[Fe]/[H]} = 1.26 \times 10^{-4}$ (Feldman \emph{et
al.}, 1992), a low-FIP element abundance enhanced by a factor of about four
respect
to typical photospheric values (Grevesse and Sauval, 1998).  The mean
temperature [$\Tm$] and the temperature spread [$\WT$] are not affected by the
assuumed Fe abundance. 

As in our previous papers (FVK09; V\'asquez \etal{} 2009; VFM10), for the
FBE to LDEM inversion we assumed the Arnaud and Raymond (1992) ionization
equilibrium calculations. In VFM10 we also performed an alternative inversion,
based on the Mazzotta \etal{} (1998) ionization equilibrium model, to evaluate
the typical uncertainty of LDEM moments due to the assumed model. The most
affected quantity was the inferred temperature spread [$\WT$], with typical
uncertainties of order 4\% or less. The least-affected result was the mean
electron temperature [$\Tm$], with uncertainties below 1\%, while the inferred
electron density [$\left< N_e^2\right>^{1/2}$] showed uncertainties of order 1\%
or less.

To quantify the uncertainty in the LDEM results due to the regularization level,
in this work we performed two separate analyses, based on reconstructions using
the mean and the minimum regularization levels obtained from the
cross-validation study. We found the largest uncertainty in the temperature
spread [$\WT$], with values in the range 3 to 9\%. The uncertainty of the
estimated electron density [$\Ne$] is in the range 2 to 5\%. The least-affected
result is the mean electron temperature [$\Tm$], with uncertainties below 2\%
everywhere.

We conclude this section with a brief discussion of the limits of validity of
the technique's results. As the SRT technique applied here does not account for
the Sun's temporal variations, rapid dynamics in the region of one voxel can
cause artifacts in neighboring ones. Such artifacts include smearing and
negative values of the reconstructed FBEs, or zero when the solution is
constrained to positive values. {These are called zero-density artifacts (ZDAs)
and are similar in nature to those described by Frazin and Janzen (2002) in the
context of white-light tomography.} The voxels belonging to ZDA regions are
excluded of the LDEM analysis.  Active regions can present particularly rapid
dynamics, and we do not analyze them here. {For all voxels with no ZDAs, we use
the inferred LDEM to forward-compute the three synthetic values of the FBE. The
synthetic and reconstructed values agree within 1\% in 77\% of the
voxels, and within 10\% in 82\% of the cases. For the analysis in Section
\ref{resultsC}, we only use the voxels where the achieved accuracy is within
1\%.}

Due to optical-depth issues in the EUV images close to the limb and the ﬁnite
extent of the EUVI field of view, we view the tomographic reconstructions in
this work as physically meaningful between heliocentric heights of 1.03 to 1.23
\rsun . As with many optical instruments, the image measured by EUVI can be
modeled as a convolution of the true solar image (as would be seen by an ideal
telescope) with the instrument point spread function (PSF). The PSF has
important consequences for the Sun's fainter structures such as coronal holes
(CHs) and emission at larger heights above the limb.  Our preliminary analysis
shows that, depending on the band, up to about 50\% of the emission seen in CHs
is due to the PSF.  Since our deconvolution procedures are not yet ready for
deployment, we do not analyze CHs here.

\section{Observational Data, Tomography Parameters, and PFSS Model}\label{OBS}

We use EUVI/STEREO A and B data taken simultaneously during Carrington rotation
(CR) 2068 (March 20 01:14 UT through 2008 April 16 08:05 UT 2008). In this
period, the spacecraft were separated by an average of 47.7\deg, which allowed
for the reconstruction to be performed with data gathered in about 24 days, a
little less than the rotational period. The data set used consists of one-hour
cadence images taken in the 2008 period between 20 March 00:00 UT and 12 April
23:59 UT. The total number of images used from each instrument and band is then
about 576. During the observational period, the two spacecraft separation
provided redundant observations over a range of about 265\deg. This resulted in
a rich dataset, which provides much information for cross-validation purposes,
as explained below.

The spherical computational grid covers the height range 1.00 to 1.25 \rsun with
25 radial, 90 latitudinal, and 180 longitudinal bins, which gives a total number
of about $4\times 10^5$ voxels, each with a uniform radial size of 0.01 \rsun\
and a uniform angular size of 2\deg\ (in both latitude and longitude). It is not
useful to constrain the tomographic problem with information taken from view
angles separated by less than the grid angular resolution. Therefore, as the Sun
rotates about 13.2\deg per 24-hour period, we time average the images in 6-hour
wide bins, so that each time-averaged image is representative of views separated
by about 3.3\deg. The total number of time-averaged images from each instrument
and band is then about 96. Due to their high spatial resolution ($1.6''$ per
pixel), {to reduce both memory load and computational time}, we spatially rebin
the images by a factor of eight, bringing the original 2048 $\times$ 2048 pixel
EUVI images down to 256 $\times$ 256. Thus the final images' pixel size is about
the same as the radial voxel dimension. Due to this spatial and temporal
binning, the statistical noise in the EUVI images is greatly reduced.

Due to the relative spacecraft positions, the 24 days of collected data implied
an angular range of 265\deg\ in which both spacecraft saw the Sun from almost
exactly the same viewpoint, although at different times.  This resulted in a
data set with 81 redundant image pairs.  These redundant images give us the
opportunity to determine the regularization parameter [$p$], by finding the
value that best predicts one set of the redundant data, \textit{i.e.} the 81 A
or B images. This is only one way of choosing validation data, and Frazin and
Janzen (2002) performed cross validation with single spacecraft data. Using the
same cross-validation procedure described in VFM10, we obtained values in the
range \textit{p} = 0.35 $\pm$ 0.15, with the difference most like being due to
the change in the spacecraft-separation angle The similar study for the
reconstructions in V\'asquez \etal{} (2009) and VFM10 gave comparable ranges,
centered in the values $p=0.9$ and $p=1.77$, respectively. The results presented
in this work correspond to $p=0.5$, using all images from both spacecraft. In
Section \ref{results} we quantify the uncertainty of the LDEM moments due to
that of the regularization parameter. Regularization parameter selection and
other uncertainty quantification will be given a comprehensive treatment in a
forthcoming publication.  

We made use of a potential field source surface (PFSS) model of the coronal
magnetic field (Altschuler \etal, 1977). The source-surface height was set at
\rss\ = 2.5 \rsun, and the lower boundary condition prescribed by the synoptic
magnetogram for CR-2068 provided by MDI/SOHO. We traced the magnetic-field lines
through the tomographic computational grid, producing the run of LDEM
distributions along each field line.  This also allowed classification of the
voxels as belonging to magnetically open or closed regions of the PFSS model.

\section{Results}\label{results}

\subsection{Analysis of the 3D Reconstructions}\label{resultsA}

\begin{figure}[ht]
\includegraphics[width=\linewidth]{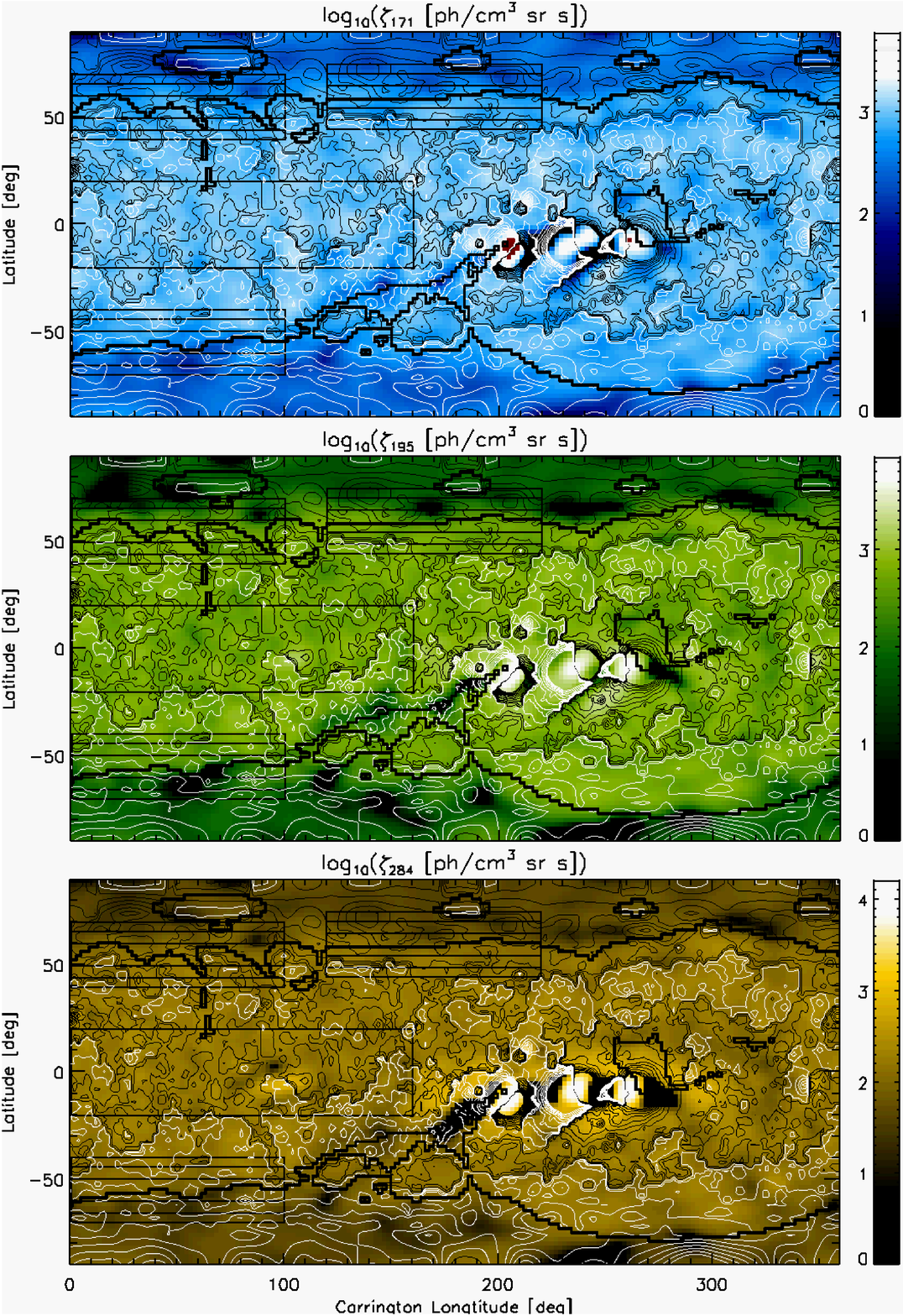}
\caption{Carrington maps of the reconstructed 3D FBEs $\zeta_{k}$ at a height of
1.035 \rsun, for the three coronal Fe bands at 171, 195, and 284 \AA. The
overplotted solid-thin curves are magnetic strength [$B$] contour levels from
the PFSSM taken at the same height, with those in white (black) representing
outward (inward) oriented magnetic field. The solid-thick black curves mark the
magnetically open/closed regions boundary. The boxes highlight the regions
analyzed in Section \ref{resultsC}.}
\label{FBE_maps_1}
\end{figure}

As an example of the 3D tomographic reconstruction of the  emissivity, Figure
\ref{FBE_maps_1} shows the Carrington maps of $\zeta_{k}$ at height 1.035 \rsun,
for the three coronal Fe bands of 171, 195, and 284 \AA.  The black voxels are
the ZDAs described in Section 2, which occupy 9\% of the reconstructed volume.
The overplotted solid-thin curves are contour levels of the PFSSM magnetic
strength [$B$], in steps of 1.0 G. The white (black) contours represent outward
(inward) oriented magnetic field. The overplotted solid-thick black curves
indicate the boundary between the magnetically open and closed regions. The
reconstructed FBEs exhibit larger values within the PFSS model's magnetically
closed regions, and lower values in the open regions. The peak emission is
located within the active region (AR) complex in the southern hemisphere and
near the Equator, in the longitudinal range [190\deg, 270\deg]. 
{The PFSS model considers latitudes up to $\pm 81.5\mdeg$, beyond which the
magnetogram was extrapolated. At the larger latitudes, the MDI data shows an
asymmetry between both hemispheres, with the northern part exhibiting a
decreasing magnetic strength beyond latitude +75\deg. This gives rise to
artificial magnetically neutral locations in the extrapolated highest latitudes
of the northern hemisphere, producing an artifact of $\beta>1$ values at all
longitudes near the North Pole (see  Figure \ref{beta_maps} and its discussion
below.}

\begin{figure}[ht]
\includegraphics[width=\linewidth]{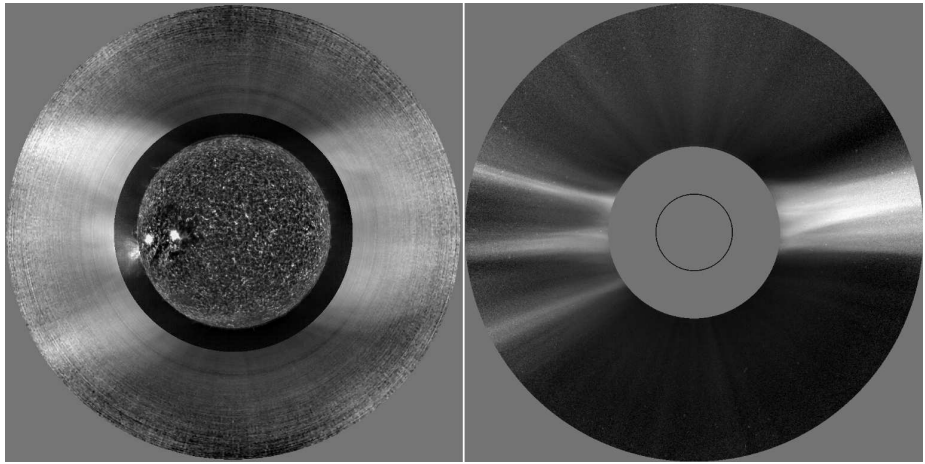}
\caption{The solar corona on 2008 March 24, taken from the solar data catalog of
the Institute for Astronomy (IfA) of the University of Hawaii
(http://alshamess.ifa.hawaii.edu). The images of the different instruments were
taken within a fourty minute lapse centered at 19:00 UT. Left: a SOHO EIT 304
\AA\ image, surrounded by a MLSO coronagraph white light image. Right: SOHO
LASCO C2 white light image.}
\label{context}
\end{figure}

{In the following discussion, we identify the magnetically closed region
as the equatorial \emph{streamer core}. The magnetically open field lines
immediately surrounding the equatorial streamer core are known as the
\emph{streamer legs}, where the O VI 1032 \AA\ intensity relative to H I 1216
\AA\ is often seen to be higher than in the core, above 1.5 \rsun, as seen for
example in Raymond et al. (1997) and Strachan et al. (2002). A geometrical
sketch of the streamer core and leg structure can be found in Figure 1 of Nerney
\& Suess (2005). The magnetically open latitudes just outside of the streamer
legs, up to about latitudes $\pm 75\mdeg$, will be referred to as the subpolar
regions. Beyond that latitude we curretly avoid analysis of results  due to the
importance of PSF contamination in CHs (deconvolution has not yet been
implemented). Figure \ref{context} displays the solar corona on 2008 March 24 at
approximately 19:00 UT. In those images the west and east limb longitudes are
about 30\deg\ and 210\deg, respectively. In the EIT image, the east limb just
hides the foot location of the easternmost AR seen between longitudes 200\deg
and 210\deg\ in the Carrington maps of Figure \ref{FBE_maps_1}. In the west
limb, the \textit{Mauna Loa Solar Observatory} (MLSO) MkIV white light image
shows the streamer belt between heliocentric heights 1.25 and 2.35 \rsun, around
longitude 30\deg. That location is surrounded by a very wide longitude range of
quiet sun corona, at least 90\deg\ wide in both the eastward and westward
directions (see Figure 1). The Large Angle and Spectrometric Coronagraph
(LASCO/SOHO) C2 image to the right shows the extended streamer structure above
2.3 \rsun.} Being a period of minimum activity, open regions are usually
confined to the higher latitudes, and characterized by a lower emissivity. The
PFSS model open/closed boundary shows an isolated low-latitude open region,
centered at longitude 265\deg\ and latitude +5\deg\ at 1.035 \rsun. In the
southern hemisphere, the open regions extend to low latitudes within the
longitude range [100\deg, 200\deg], reaching a maximum latitude of $-10\mdeg$\
around longitude 200\deg.
 
\begin{figure}[ht]
\includegraphics[width=\linewidth]{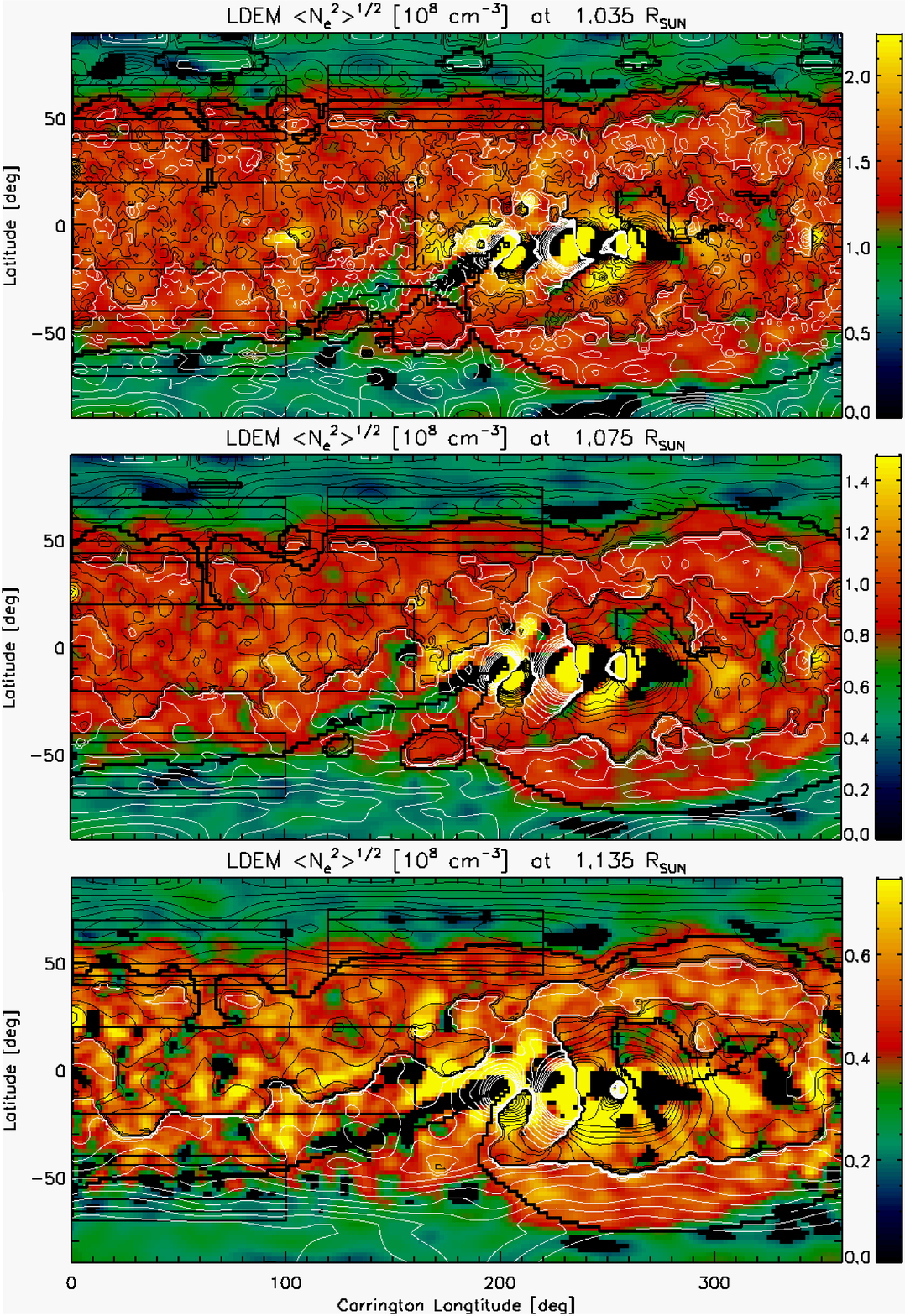}
\caption{Carrington maps of the LDEM $\left< N_e^2\right>^{1/2}$ at heights
1.035, 1.075 and 1.135 \rsun, from top to bottom. Solid-thin curves are magnetic
strength [$B$] contour levels from the PFSSM taken at the same height, with
those in white(black) representing outward(inward) oriented magnetic field. The
solid-thick black curves mark the magnetically open/closed regions boundary.}
\label{LDEM_Ne_maps}
\end{figure}

\begin{figure}[ht]
\includegraphics[width=\linewidth]{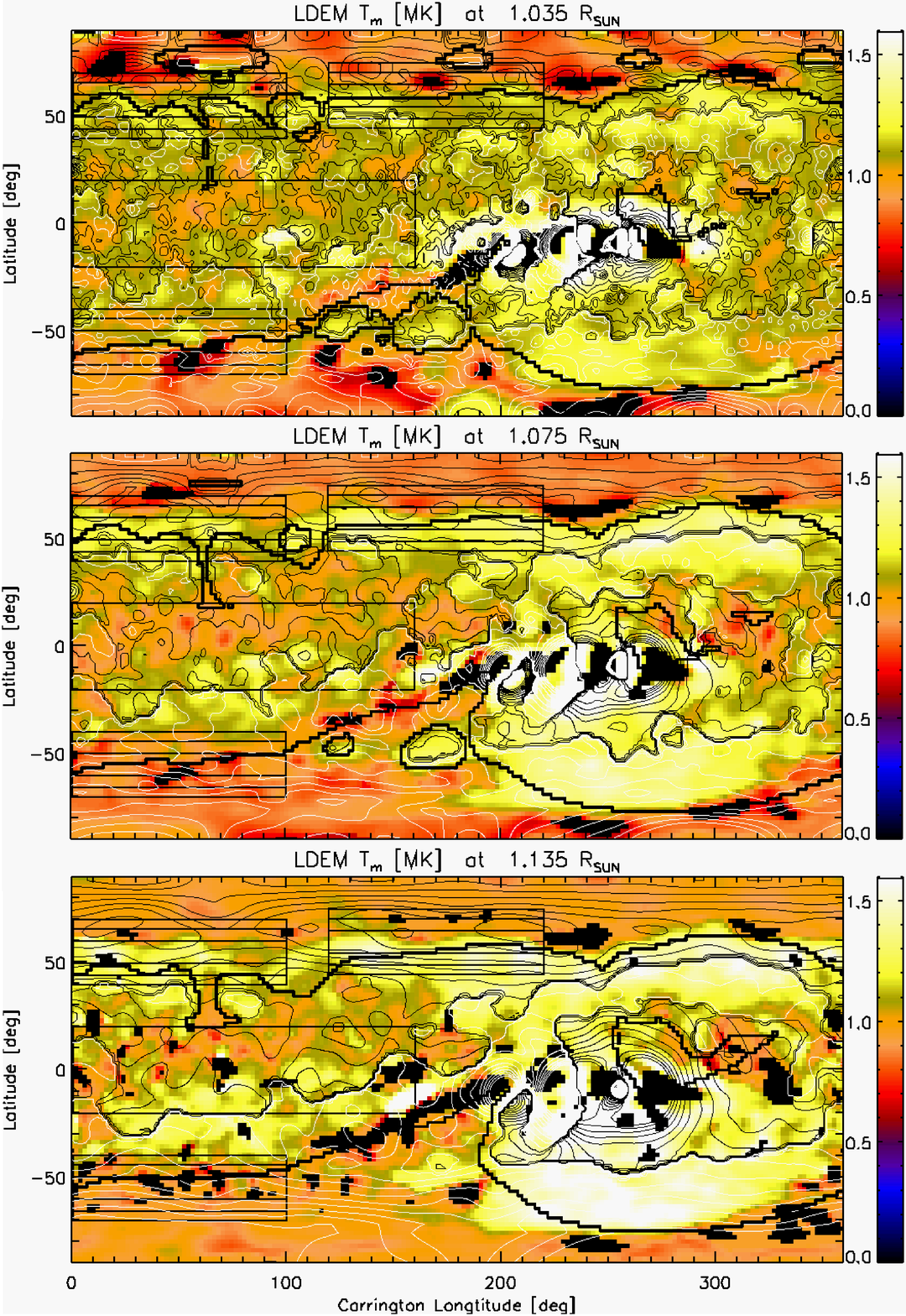}
\caption{Carrington maps of the LDEM $\Tm$ at heights 1.035, 1.075 and 1.135
\rsun, from top to bottom. Solid-thin curves are magnetic strength [$B$] contour
levels from the PFSSM taken at the same height, with those in white (black)
representing outward (inward) oriented magnetic field. The solid-thick black
curves mark the magnetically open/closed regions boundary.}
\label{LDEM_Tm_maps}
\end{figure}

\begin{figure}[ht]
\includegraphics[width=\linewidth]{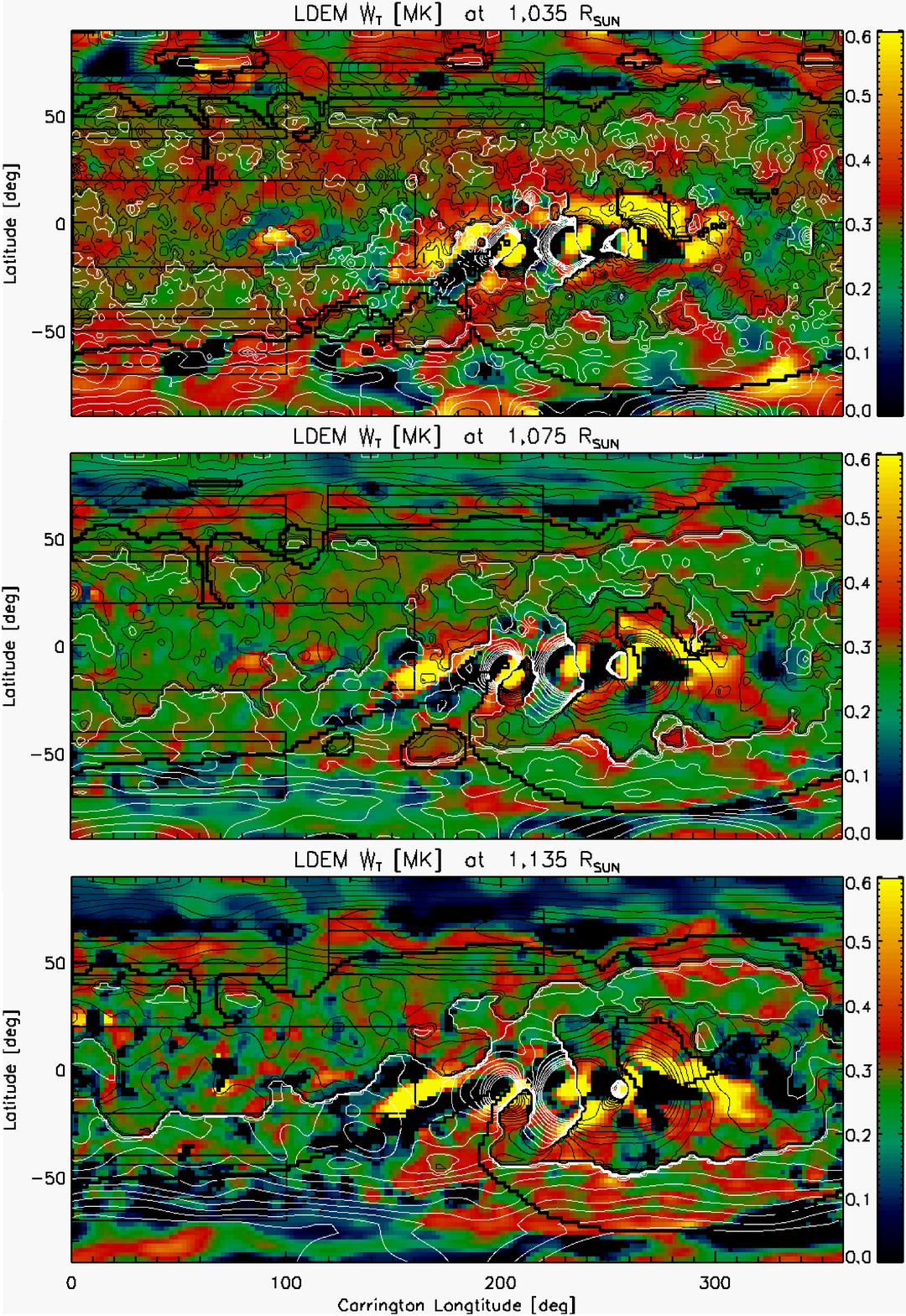}
\caption{Carrington maps of the LDEM $W_T$ at heights 1.035, 1.075 and 1.135
\rsun, from top to bottom. Solid-thin curves are magnetic strength [$B$] contour
levels from the PFSSM taken at the same height, with those in white (black)
representing outward (inward) oriented magnetic field. The solid-thick black
curves mark the magnetically open/closed regions boundary.}
\label{LDEM_WT_maps}
\end{figure}

Figure \ref{LDEM_Ne_maps} shows the Carrington maps of the 3D distribution of
the electron density [$\left< N_e^2\right>_i^{1/2}$, in $10^8\, {\rm cm}^{-3}$
units], at heights 1.035, 1.075, and 1.135 \rsun. For the same heights, Figures
\ref{LDEM_Tm_maps} and \ref{LDEM_WT_maps} show the Carrington maps of the mean
temperature [$T_{m,i}$, MK] and the temperature spread [$W_{T,i}$, MK],
respectively. In all these Figures, we overplot PFSSM magnetic strength
[$B$] contour levels (solid-thin black and white curves), as well as the
magnetically open/closed region boundaries (solid-thick black curves). The black
voxels in the different maps correspond to undetermined LDEM locations due to
the presence of a ZDA for at least one of the bands, which represent 9\% of the
total number of voxels.

Figure \ref{LDEM_Ne_maps} shows that the largest densities are found in the ARs.
In order to highlight the quiescent structure, we threshold the maximum density
displayed at each height, as indicated by the respective color scales. The peak
densities in the AR complex are 9.0, 6.3, and 3.9 $\times 10^8\, {\rm cm}^{-3}$
at 1.035, 1.075, and 1.135 \rsun, respectively. 

At all heights, the PFSSM closed region is populated by densities clearly
enhanced with respect to the open regions, consistent with its identification as
the streamer core. A notable characteristic of these maps is that, at all
heights, the PFSSM open/closed boundary very accurately demarcates the location
of the transition between streamer to subpolar density levels (red to green).
This good morphological match between the structures derived from the
tomographic and the PFSSM analyses, is similar to what we found for CR-2077
(VFM10), compared to which the topology found here is more complex due to a
relatively higher level of activity. At all heights, the magnetically open
subpolar regions are characterized by densities of order half of those typically
found within the quiescent streamer core, which are typically in the range
$1.5-0.5\times10^8\, {\rm cm}^{-3}$ in the height range we study.

Figure \ref{LDEM_Tm_maps} shows that the largest temperatures are found in the
ARs. In order to highlight the quiescent structure, we threshold the maximum
temperature displayed at each height, as indicated by the respective color
scales. The peak $\Tm$ values in the AR complex are 2.5, 2.4, and 2.9 MK at
1.035, 1,075, and 1.135 \rsun, respectively. The Carrington maps show that the
most quiet longitudes of the streamer core, away from the AR complex, are
characterized by lower temperatures [$\Tm$] near the Equator than at higher
latitudes. These higher temperature regions can typically reach $\Tm = 1.4$ MK,
being 40\% larger than average streamer core equatorial values. The overplotted
magnetic-field strength contour levels reveal that all high $\Tm$ (yellow) areas
are located along and around polarity inversion lines, as well as close to the
open/closed boundary. Note also that, in most cases, these higher-temperature
regions do not extend into the magnetically open parts, and the open/closed
boundary generally matches their high latitude limit. The characteristics of the
coronal temperature structure described in this paragraph are common to the
DEMT results of the CR-2077 period (VFM10).

Figure \ref{LDEM_WT_maps} shows that the largest values of the LDEM electron
temperature  spread [$\WT$] are found in the ARs, where the peak value is around
0.85 MK at all heights. In order to highlight the quiescent structure, we
threshold the maximum displayed value at each height, as indicated by the
respective color scales. Figure \ref{LDEM_WT_maps} shows that the distribution
of  $\WT$ is  quite complex, not clearly correlated with the other two moments,
except for the AR complex, outside of which the temperature spread values are
typically in the range 0.1 to 0.4 MK. Unlike for the other two moments, the
magnetically open/closed boundary is not characterized by a clear change of
temperature spread. The lowest values are seen in the open regions, but
generally away of the open/closed boundary, at somewhat higher latitudes. This
has been found more clearly in our analysis of the period CR-2077 (VFM10). In
that period, characterized by a much simpler coronal structure, we found that
the change from higher to lower $\WT$ values systematically occurred some
degrees in latitude outside the magnetically open/closed boundary.

\begin{figure}[ht]
\includegraphics[width=\linewidth]{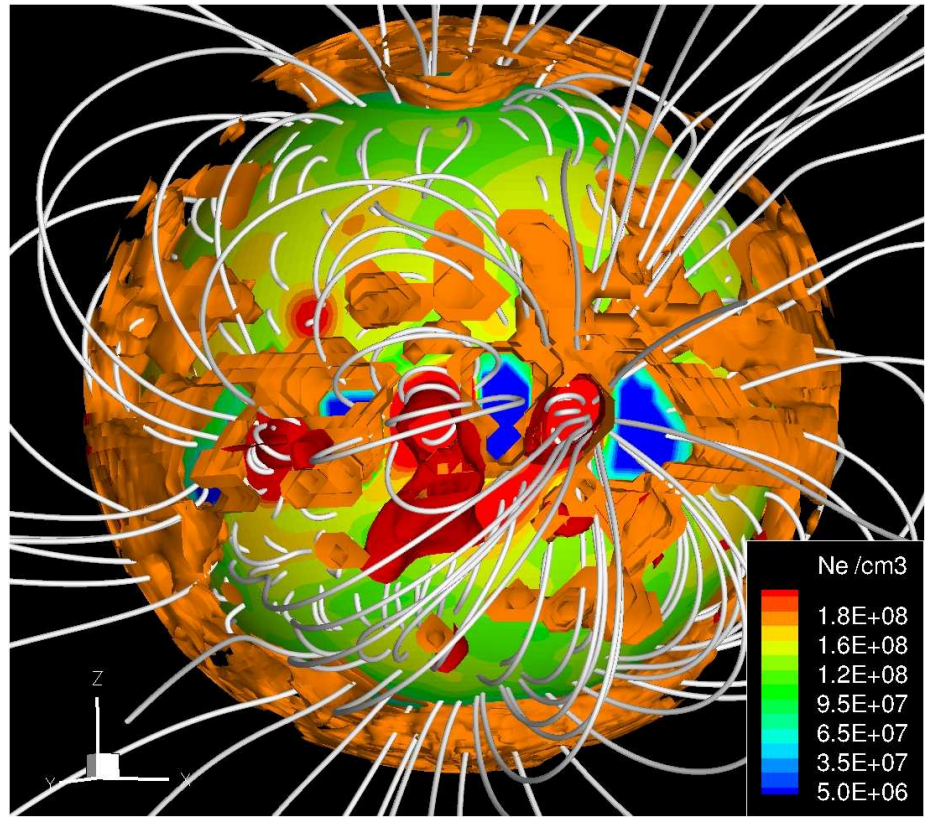}
\caption{A 3D view of the CR-2068 PFSS model. Some representative open and
closed magnetic-field lines are drawn in white. The red and orange regions are
LDEM $\Tm$ isosurfaces of 2 and 1 MK, respectively.  The inner spherical surface
shows the LDEM $\Ne$ at 1.04 \rsun.}
\label{3Dview}
\end{figure}

\begin{figure}[ht]
\includegraphics[width=\linewidth]{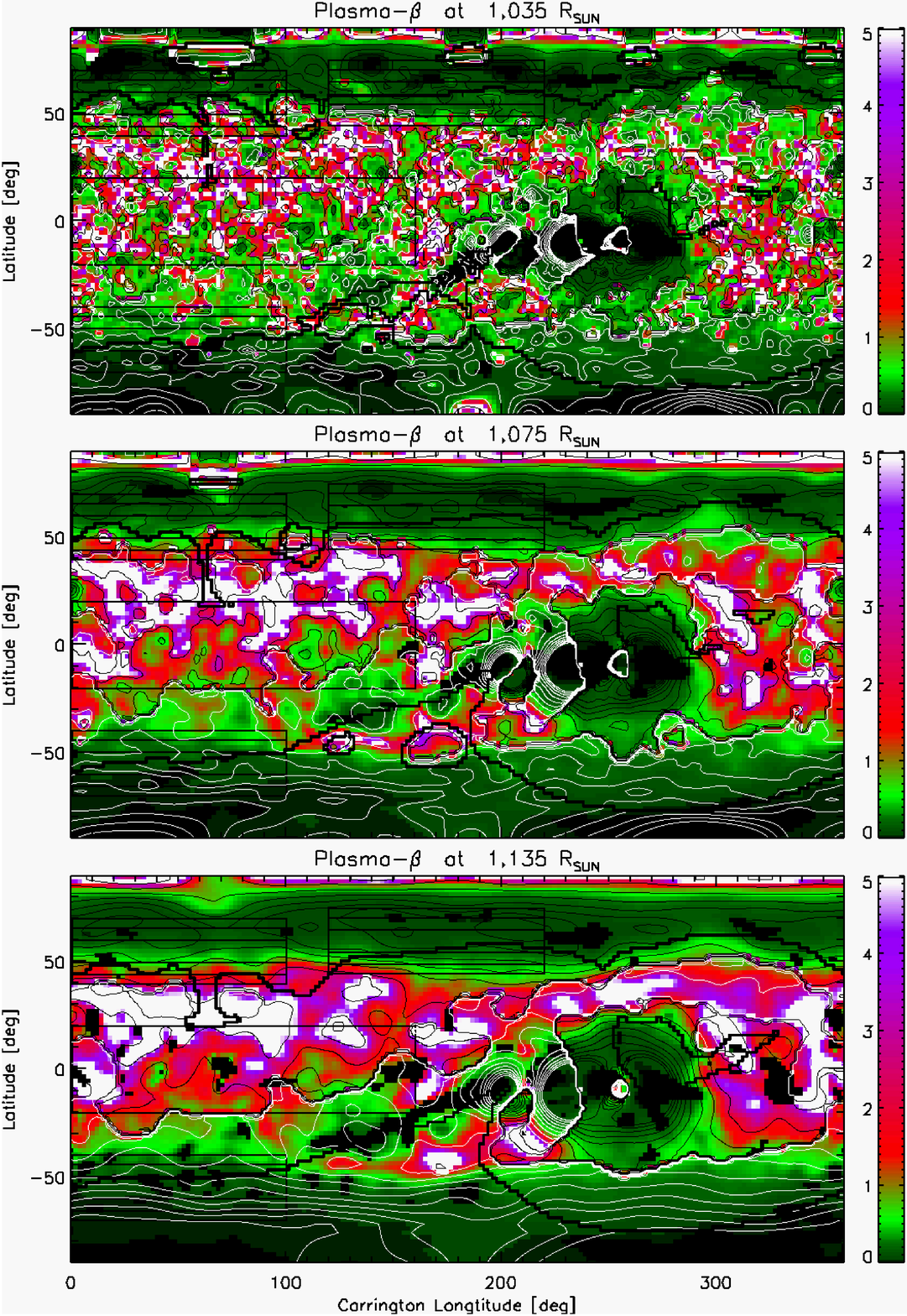}
\caption{Carrington maps of the plasma $\beta$ at heights
1.035, 1.075 and 1.135 \rsun, from top to bottom. Solid-thin curves are magnetic
strength [$B$] contour levels from the PFSSM taken at the same height, with those
in white (black) representing outward (inward) oriented magnetic field. The
solid-thick black curves mark the magnetically open/closed regions boundary.}
\label{beta_maps}
\end{figure}

{Figure \ref{3Dview} shows a 3D view of the CR-2068 PFSS model. Some
representative open and closed magnetic-field lines are drawn in white. The red
and orange regions are LDEM $\Tm$ isosurfaces of 2 and 1 MK, respectively.  The
inner spherical surface shows the LDEM $\Ne$ at 1.04 \rsun, with the displayed
color scale. The 2 MK region corresponds to the AR complex. Figure
\ref{LDEM_Tm_maps} shows that at all heights in the northern hemisphere, the
magnetically open/closed boundary and its surrounding latitudes show
$\Tm>1\,{\rm MK}$. On the magnetically open side of the boundary, the 1 MK level
(orange voxels) is achieved at latitudes right outside the open/closed boundary.
On the magnetically closed side of the boundary, the 1 MK level  is achieved at
latitudes considerably lower than those of the open/closed boundary.
Consistently, the orange isosurface of the 3D view leaves then a wide ``empty''
region just inside of the magnetically open/closed boundary, and the border of
the orange isosurface in the northern polar region  indicates  the open/closed
boundary quite accurately.}

With the tomographic LDEM moments $\momNe$ and $\Tm$, as measures of the
electron density and temperature, respectively, and the PFSSM magnetic strength
[$B$], we estimate the plasma $\beta=16\pi\momNe k_B \Tm / B^2$, where we have
approximated the total thermal pressure as $p\approx 2 \Ne k_B \Te$. Figure
\ref{beta_maps} shows Carrington maps of $\beta$ at heights 1.035, 1.075, and
1.135 \rsun, thresholded at a maximum value of five. It is interesting to note
how $\beta>1$ values are typical of the streamer core region, except for the ARs
complex where the magnetic strength is very high. The largest $\beta$ values
within the streamer core are of order 5 to 10, corresponding to the lowest
magnetic-strength values near polarity inversion zones. Throughout the
magnetically open regions $\beta<1$, except for the artifacts in the northern CH
region mentioned above.

\subsection{Analysis of Results for Quiet Sun Open and Closed Magnetic
Structures}\label{resultsC}

We analyze here the physical properties of quiet-Sun regions as derived with the
LDEM analysis, both in the streamer core and in the  surrounding magnetically
open regions. In each panel of Figure \ref{FBE_maps_1}, the four black boxes
highlight the quiet-Sun regions selected for the analysis. In the equatorial
region we selected the streamer core between longitudes 0\deg\ and 160\deg, and
latitudes -20\deg\ to +20\deg, a region that we name here SC. The other
selected regions surround the open/closed boundary, and we name them here:
\R{1} (southern hemisphere, between longitudes 0\deg\ and 100\deg, and latitudes
-70\deg\ to -40\deg), \R{2} (northern hemisphere, between longitudes 0\deg\ and
100\deg, and latitudes 40\deg\ to 70\deg), and \R{3} (northern hemisphere,
between longitudes 120\deg\ and 220\deg, and latitudes 45\deg\ to 75\deg). Note
that in each of these regions the open/closed boundary occupies the intermediate
latitude range. The horizontal black lines within each of these three regions
delimit subregions that are discussed below.

\begin{figure}[ht]
\includegraphics[width=\linewidth]{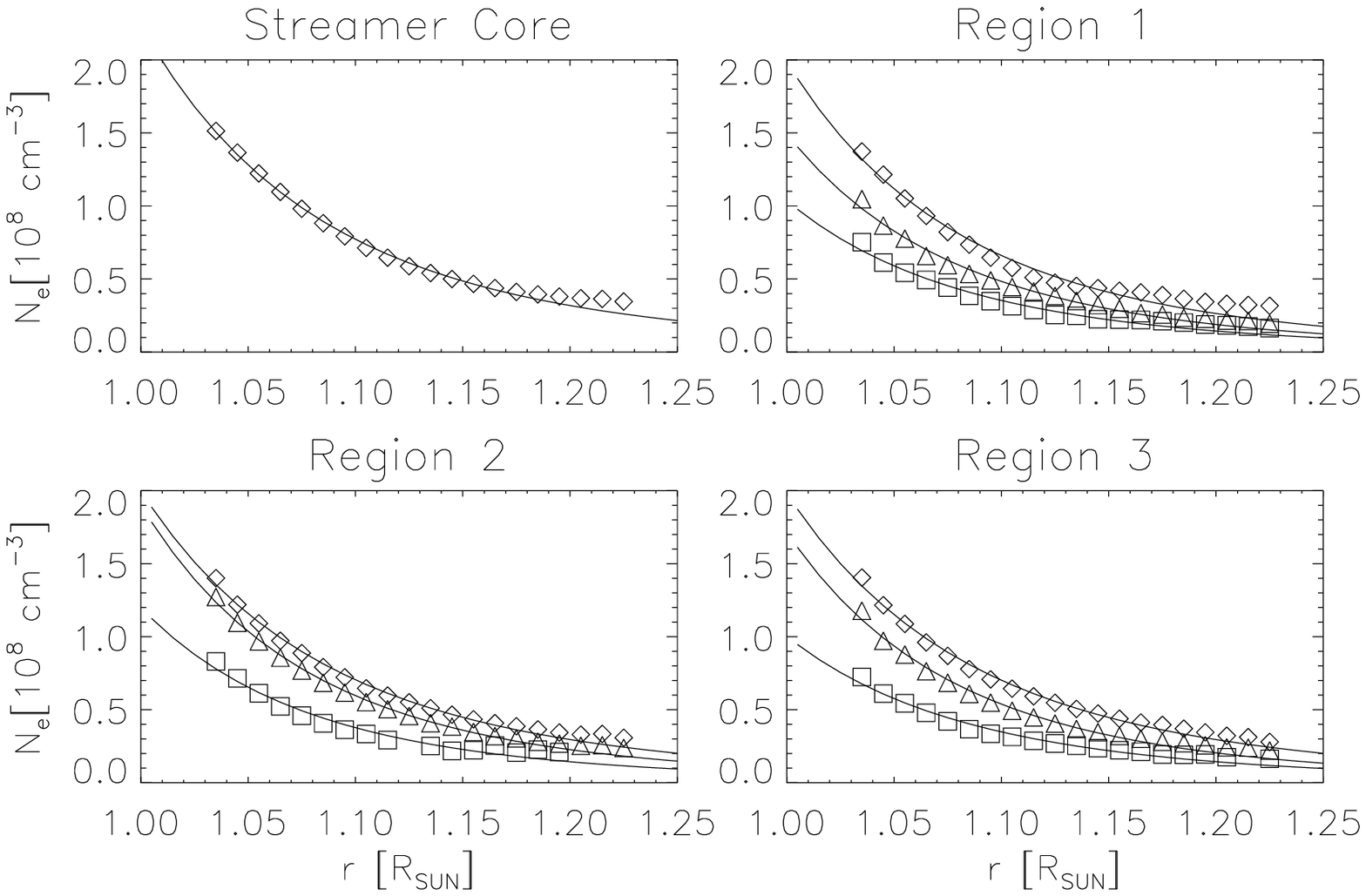}
\caption{Average dependence with height of the LDEM electron density [$N_e(r)$]
in the ten regions SC and \R{i}-{O,B,C} of Table 1. The LDEM data is indicated
by symbols, while the solid curves show the best hydrostatic fits given by
Equation (1). In regions \R{i}, the diamonds indicate the closed part of the
data inside the streamer, the triangles are the data around the open/closed
boundary, and the squares are the data in the open subpolar parts.}
\label{tomo_height_Ne}
\end{figure}

\begin{figure}[ht]
\includegraphics[width=\linewidth]{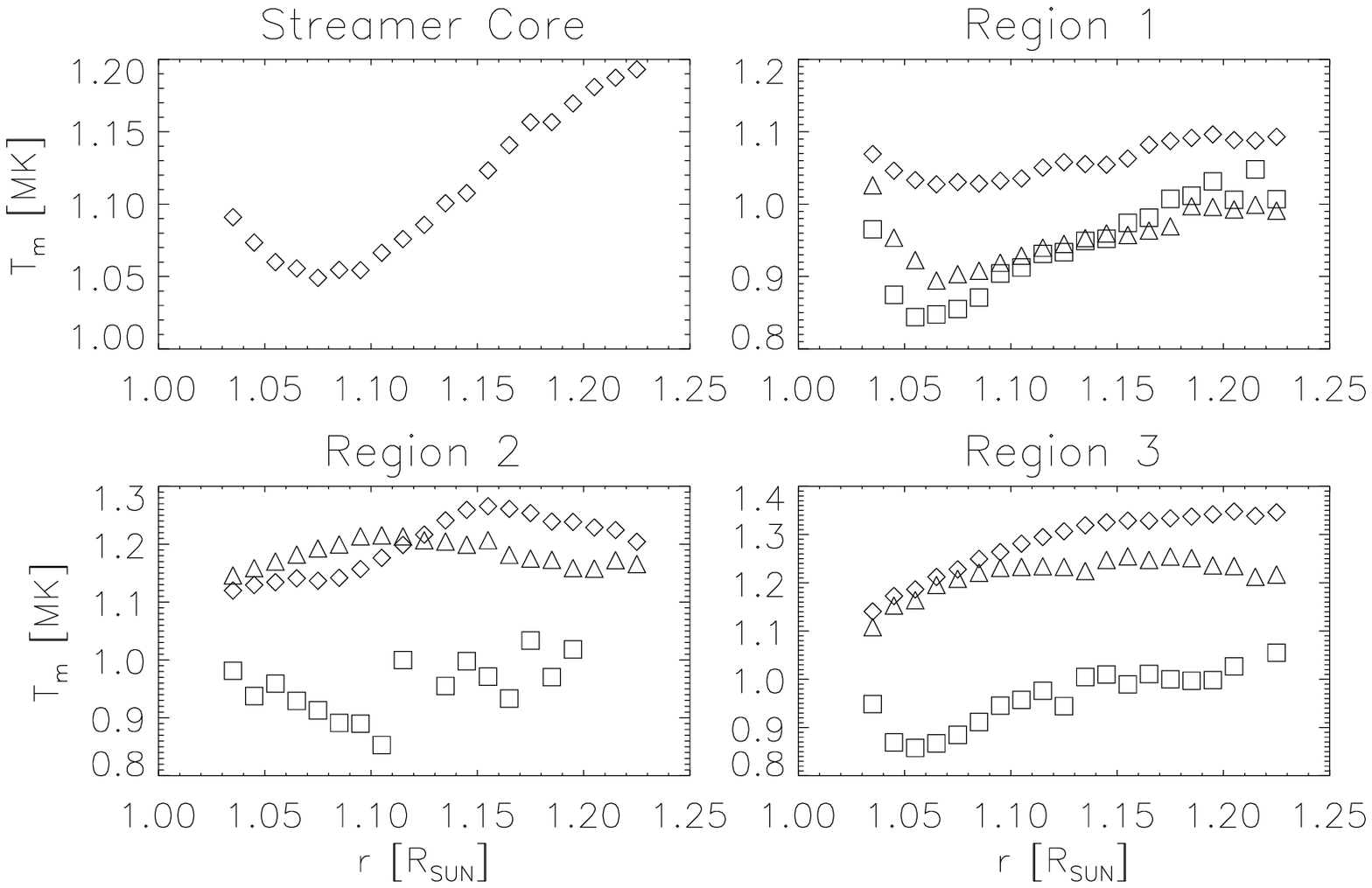} \caption{Average
dependence with height of the LDEM electron mean temperature [$\Tm(r)$] in the
ten regions SC and \R{i}-{O,B,C} of Table 1. The LDEM data is indicated by
symbols, while the solid curves show the best hydrostatic fits given by Equation
(1). In regions \R{i}, the diamonds indicate the closed part of the data inside
the streamer, the triangles are the data around the open/closed boundary, and
the squares are the data in the open subpolar parts.}
\label{tomo_height_Tm}
\end{figure}

\begin{figure}[ht]
\includegraphics[width=\linewidth]{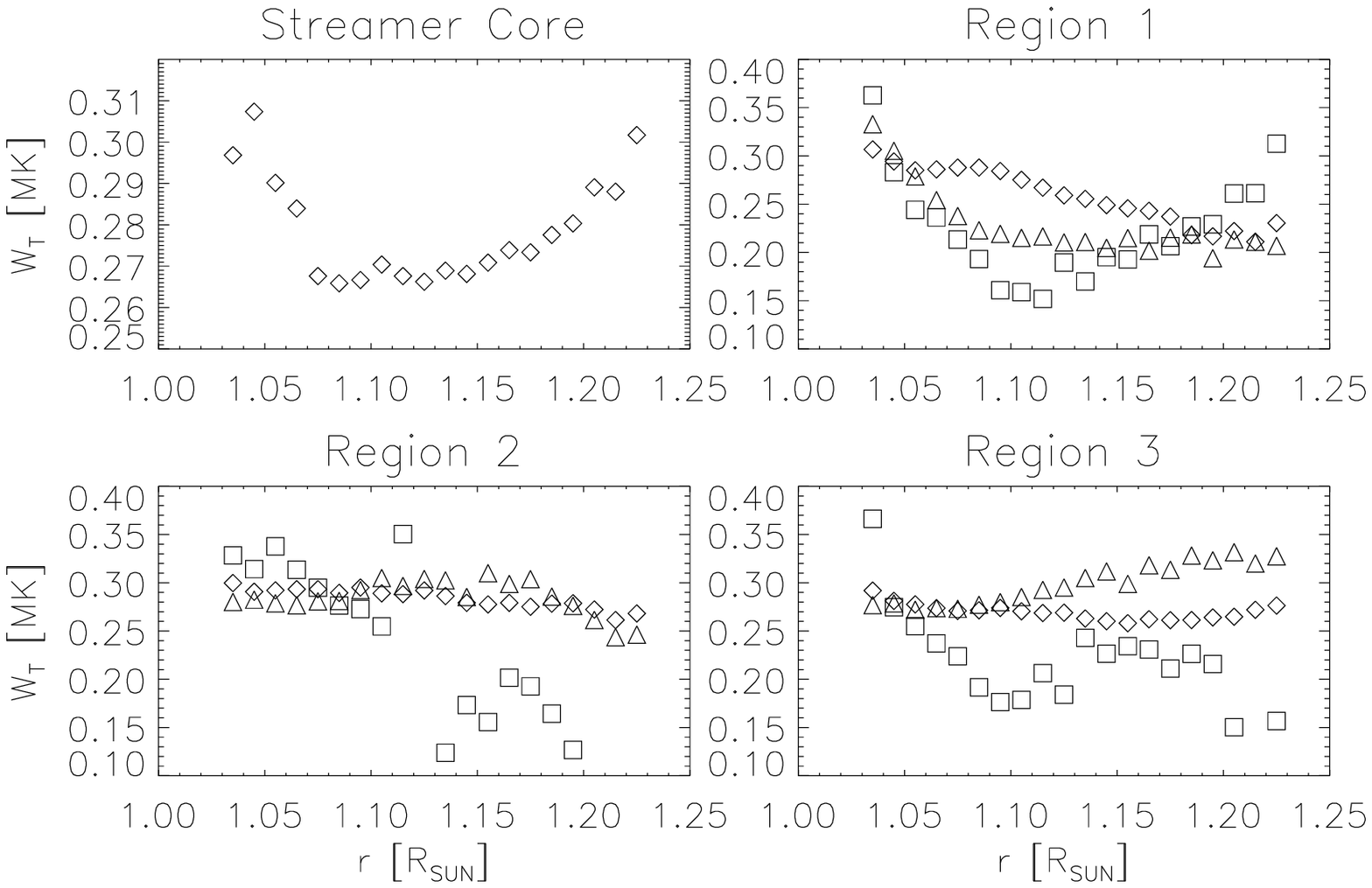} \caption{Average
dependence with height of the LDEM electron temperature spread [$\WT(r)$] in the
ten regions SC and \R{i}-{O,B,C} of Table 1. The LDEM data is indicated by
symbols, while the solid curves show the best hydrostatic fits given by Equation
(1). In regions \R{i}, the diamonds indicate the closed part of the data inside
the streamer, the triangles are the data around the open/closed boundary, and
the squares are the data in the open subpolar parts.}
\label{tomo_height_WT}
\end{figure}

In order to characterize the plasma properties in the streamer/subpolar
transition, we analyze the LDEM results as a function of height when averaged
over the regions \R{i} (\textit{i}=1,2,3) and the streamer core. Within each of
the
regions \R{i}, we have divided our analysis in three different latitude
sub-ranges, that we name here as \R{i}-C, \R{i}-B and \R{i}-O. Sub-regions
\R{i}-C consist of the lowest 4\deg\ in latitude of each region, sampling data
representative of their magnetically closed sector, inside the streamer core.
Sub-regions \R{i}-O consist of the largest 4\deg\ in latitude of each region,
sampling data representative of their magnetically open part, outside the
streamer core. Sub-regions \R{i}-B consist of the middle 10\deg\ in latitude of
each region, sampling data representative of the open/closed boundary zone,
mixing data from regions immediately inside and outside the open/closed
boundary. For each of these regions, Figures \ref{tomo_height_Ne} to
\ref{tomo_height_WT} show the average dependence with height of the LDEM
electron density [$N_e(r)$, which we will denote $\momNe$ hereafter], the mean
electron temperature [$\Tm(r)$], and the electron temperature spread [$\WT(r)$].
In
regions \R{i} we use diamonds to display data from subregions \R{i}-C, triangles
for \R{i}-B, and squares for \R{i}-O. 

{The solid curves in the density plots show the unweighted least-squares fit to
the tomographic data in the height range 1.035 to 1.195 \rsun, of the form
\begin{equation}\label{static}
N_e(r)=N_{e0} \ {\rm exp}\left[ -\, (h/\l)\, / \, (r/\mrSun)\right]
\end{equation}
where $r$ is the heliocentric height, $h=r-\mrSun$, $\l$ is the density scale
height, and $N_{e0}$ is the electron density at $r=1\mrSun$.} Equation
(\ref{static}) represents the hydrostatic solution for a plasma with a uniform
temperature.  In both the streamer legs and the subpolar open regions, at the
very low height range that we analyze, the the inertial effects of the bulk
outflow
velocity can be safely neglected. In a fully ionized corona, with a helium
abundance $a\equiv N({\rm He})/N({\rm H})$, the mean atomic weight per electron
is $\mu=(1+4a)/(1+2a)$, in terms of which the plasma mass density is $\rho=\mu
m_H N_e$. We set $a=0.08$, as in the CHIANTI coronal abundances set (Feldman
\etal, 1992) we used to compute the DEM kernels. The fitted scale $\l$
determines
$\Tfit$ and is given by $\l = k_B \Tfit / \left(\mu m_H\gSun\right)$, where
$\gSun\equiv G \MSun/\mrSun^2$, and
\begin{equation}\label{Tfit}
\Tfit  \equiv \Te + \TH / (1+2a) + a\ \THe / (1+2a).
\end{equation}
In case of isothermality among the different species, one obtains
$\Tfit=\Te/\bb$, with $\bb\equiv(1+2a)/(2+3a)\approx 0.52$. 

In Table \ref{tab1}, we show the statistics of the hydrostatic fit and LDEM for
the selected Regions (rows). In the different columns we tabulate the fit's
basal density $N_{e0}$, scale height $\l$, and electron temperature $\bb\Tfit$
derived from Equation (\ref{Tfit}) assuming species are isothermal. We also give
the average $\left<\Tm\right>$ and variance $\s2Tm$ of $\Tm(r)$ in the analyzed
height range, the temperature difference $\dT\equiv\bb\Tfit-\left<\Tm\right>$,
and the average $\left<\WT\right>$ and and variance $\ssWT$ of $\WT(r)$ in the
analyzed height range.

\begin{table}
\begin{tabular}{lcccccccc}
\hline
Region & $N_{e0}$ & $\l$ & $\bb\Tfit$ &
$\left<\Tm\right>$ & $\frac{\sqrt{\s2Tm}}{\left<\Tm\right>}$ &
$\frac{\dT}{\left<\Tm\right>}$  & $\left< W_T \right>$ &
$\frac{\sqrt{\ssWT}}{\left<W_T\right>}$\\
& [$10^8{\rm cm^{-3}}$] & [\rsun] &[MK] & [MK] & & & &\\ 
\hline
SC        &  2.23& 0.0855& 1.16& 1.10& 0.04&$+$0.06& 0.28& 0.04\\
\hline
\R{1}-C & 1.99& 0.0824& 1.12& 1.06& 0.02&$+$0.06& 0.26& 0.10\\
\R{1}-B & 1.49& 0.0805& 1.10& 0.95& 0.04&$+$0.15& 0.23& 0.16\\
\R{1}-O & 1.04& 0.0844& 1.15& 0.93& 0.06&$+$0.23& 0.21& 0.23\\
\hline
\R{2}-C  & 2.00& 0.0873& 1.19& 1.19& 0.04&$-$0.01& 0.29& 0.03\\
\R{2}-B  & 1.90& 0.0784& 1.07& 1.19& 0.02&$-$0.10& 0.29& 0.04\\
\R{2}-O  & 1.20& 0.0786& 1.07& 0.95& 0.05&$+$0.12& 0.24& 0.31\\
\hline
\R{3}-C &  1.98& 0.0875& 1.19& 1.27& 0.05&$-$0.06& 0.27& 0.03\\
\R{3}-B &  1.72& 0.0784& 1.07& 1.22& 0.03&$-$0.12& 0.29& 0.06\\
\R{3}-O &  1.00& 0.0859& 1.17& 0.95& 0.06&$+$0.23& 0.23& 0.19\\
\hline
\end{tabular}
\caption{\footnotesize Statistics of the hydrostatic fit and LDEM analysis as a
function of height for the selected regions, using regularization parameter
$p=0.5$.\vspace{-1mm}}
\label{tab1}
\end{table}

The hydrostatic fits are a quite accurate description of the observed dependence
with height in each region, with squared correlation coefficient values in the
range 0.963 to 0.997. We note also that, in all cases, the variability of the
LDEM electron mean temperature, measured by $\sqrt{\s2Tm}/\left<\Tm\right>$, is
6\% or lower, {so each region exhibits a quite uniform temperature, as assumed
by the hydrostatic fits (though there can be temperature variations on large
spatial scales as seen in Figure \ref{LDEM_Tm_maps}).}

In the closed regions, the match between $\bb\Tfit$ and $\left<\Tm\right>$ is
best, being within 6\% or better in the streamer core and the three \R{i}-C
regions. This is consistent with a plasma regime close to isothermal hydrostatic
equilibrium throughout the whole streamer core region. Note that in the three
\R{i}-O regions, the difference $\dT$ is clearly larger than in the \R{i}-C
regions, reaching values of order 23\%. This is an indication that, in open
regions immediately surrounding the streamer, the plasma departs from the
isothermal hydrostatic regime. Consistently with this picture, within each
region the \R{i}-B sub-region, which mixes plasma from both open and closed
regions, shows intermediate degrees of agreement between both temperatures. 

Where the value of $\dT$ is larger, one possibility is that other pressure
mechanisms are acting, modifying the scale height respect to that due to the
thermal pressure only. If so, the excellent goodness-of-fit everywhere, as
measured by $\RR$, suggests that such other processes must be linear in the
plasma density. Another possibility is that the isothermality among the
different species is not met. Equation (\ref{Tfit}) implies that if the ions are
hotter than the electrons then $\bb\Tfit>\Te$. Hence,  taking the LDEM $\Tm$ as
a measure of the true electron temperature, the fact that in all \R{i}-O regions
we find $\dT=\bb\Tfit - \left<\Tm \right> > 0$ implies that the ions are hotter
than the electrons in the open regions. {Replacing $\Te=\left<\Tm\right>$ in
Equation (\ref{Tfit}), we solve for $\TH$, and obtain
\begin{equation}
\frac{\TH}{\aTm} = \left( \frac{\dT}{\left<\Tm\right>} + \frac{1+a}{2+3a}
\right) \, \left( \frac{2+3a}{1+a\, \THe/\TH} \right)  
\, \approx 1 + 2 \, \frac{\dT}{\left<\Tm\right>}\, ,
\end{equation}
{where the last approximation is valid for negligible coronal He abundance, and
clearly shows that if $\dT/\aTm>0$ then $\TH>\aTm$.} Assuming
$\THe=\TH\equiv\Tions$, the maximum values $\dT/\left<\Tm\right>=+0.23$ we found
in regions \R{1}-O and \R{3}-O, imply $\Tions/\aTm\approx 1.48$. In the
quiet-Sun
regions SC and \R{i}-C, within the magnetically closed streamer core,  we
found values $\dT/\left<\Tm\right>=\pm0.06$, which implies a temperature ratio
in the range $\Tions/\aTm\approx 1.0\pm0.12$.} 

The analysis summarized in Table \ref{tab1} shows the average dependence with
height of the electron density for each selected region, not considering the
individual magnetic-field lines. In order to perform an analysis more consistent
with the magnetic geometry of the PFSS model, we traced individual field lines
through the tomographic computational grid. For each field line $i$, we
determined the dependences of the LDEM $\Ne^{(i)}(r)$ and $\Tm^{(i)}(r)$, in the
height range 1.03 to 1.2 \rsun. For every density profile $\Ne^{(i)}(r)$, we
performed a least-squares hydrostatic fit of the form given by Equation
(\ref{static}). The hydrostatic-fit temperature of each line [$\Tfit^{(i)}$] was
then compared with the respective LDEM electron temperature averaged along the
line, $\left<\Tm\right>^{(i)}$. 

\begin{figure}[ht]
\includegraphics[width=\linewidth]{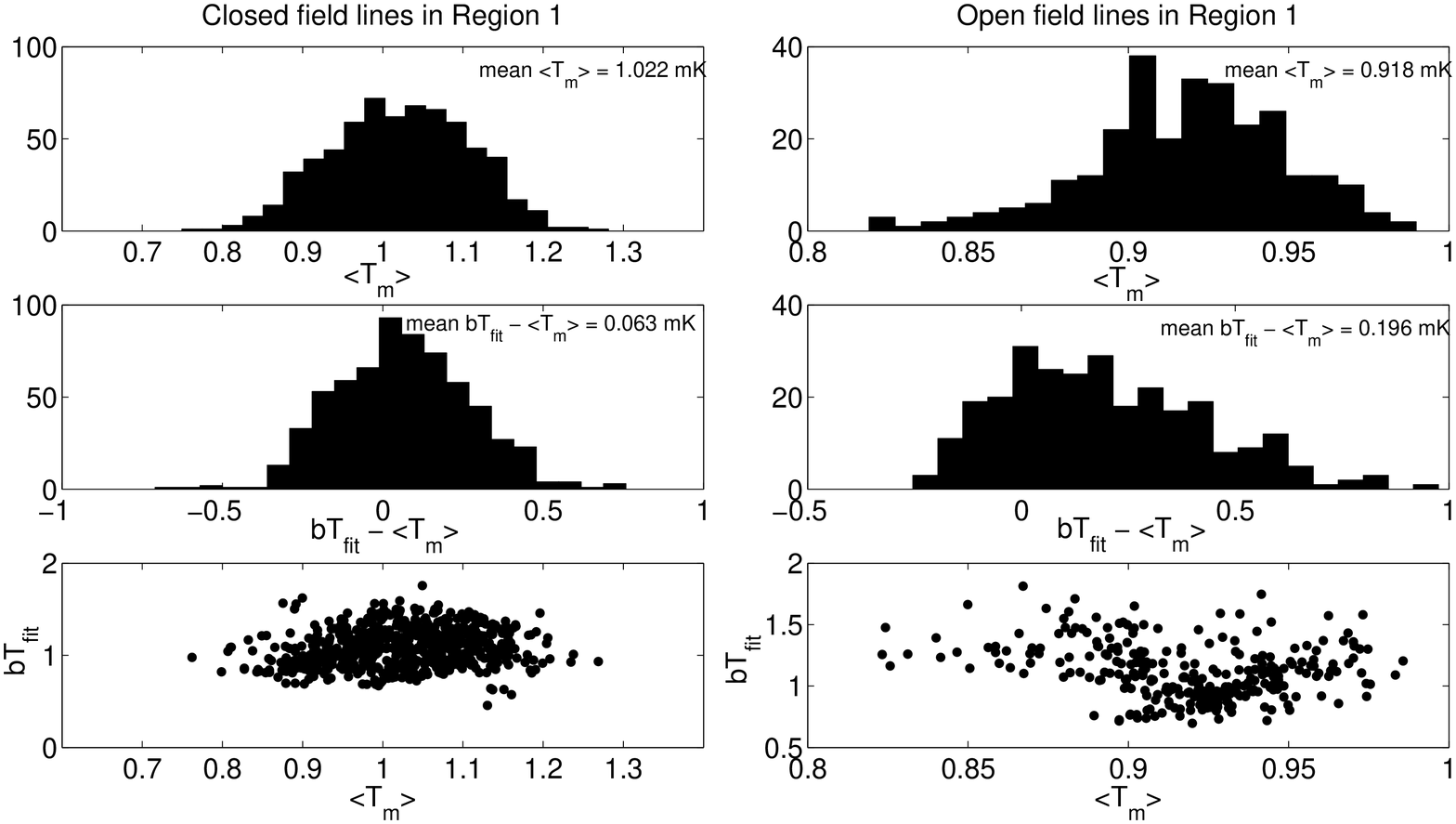}
\caption{Statistical analysis of the result of tracing individual closed (left
panels) and open (right panels) field lines $i$ in Region \R{1}. Top: histograms
of $\left<\Tm\right>^{(i)}$. Middle: histograms of $\bb\Tfit^{(i)}$. Bottom:
scatter plots of $\bb\Tfit^{(i)}$ \textit{versus} $\left<\Tm\right>^{(i)}$.}
\label{linetrace_R1}
\end{figure}

{Figure \ref{linetrace_R1} shows a statistical analysis of the results for the
closed (left) and open (right) field lines in region \R{1}. The top panels show
the histograms of  $\left<\Tm\right>^{(i)}$ along each field line. The middle
panels show the histograms of the difference
$\dT^{(i)}\equiv\bb\Tfit^{(i)}-\left<\Tm\right>^{(i)}$ for each field line. The
bottom panels shows the scatter plots of $\bb\Tfit^{(i)}$ \textit{versus}
$\left<\Tm\right>^{(i)}$. The top panels show that the two populations have
clearly different mean values, with the closed region one being $11\%$ larger.
The middle panels show a mean fractional difference
$\dT^{(i)}/\left<\Tm\right>^{(i)}$ of about $+6\%$ for the closed field lines,
and $+21\%$ for the open field lines. These numbers match very accurately those
of $\dT/\left<\Tm\right>$ in Table \ref{tab1} for \R{1}-C and \R{1}-O, which are
$+6\%$ and $+23\%$, respectively.}

\begin{figure}[ht]
\includegraphics[width=\linewidth]{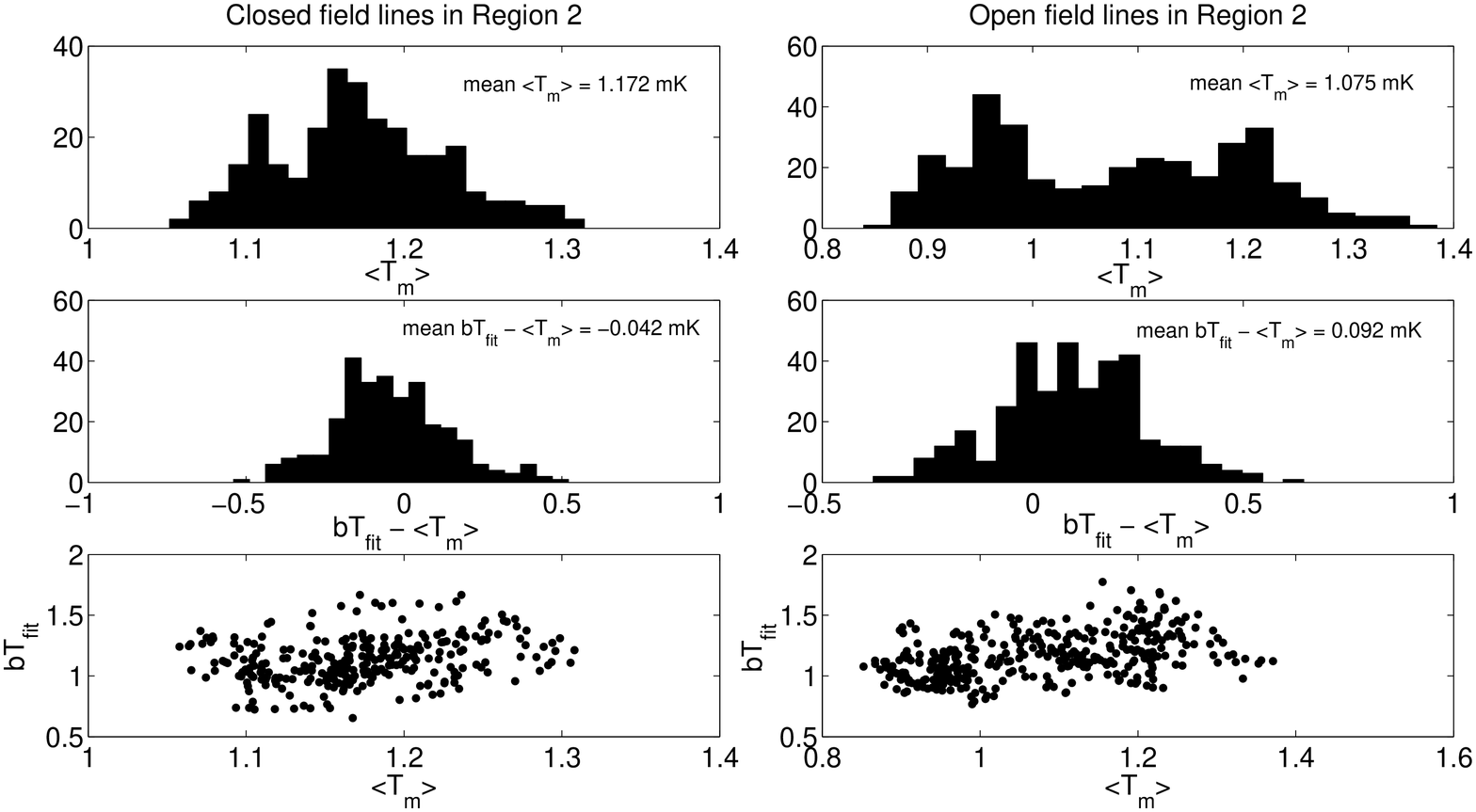}
\caption{Statistical analysis of the result of tracing individual closed (left
panels) and open (right panels) field lines $i$ in Region \R{2}. Top: histograms
of $\left<\Tm\right>^{(i)}$. Middle: histograms of $\bb\Tfit^{(i)}$. Bottom:
scatter plots of $\bb\Tfit^{(i)}$ \textit{versus} $\left<\Tm\right>^{(i)}$.}
\label{linetrace_R2}
\end{figure}

\begin{figure}[ht]
\includegraphics[width=\linewidth]{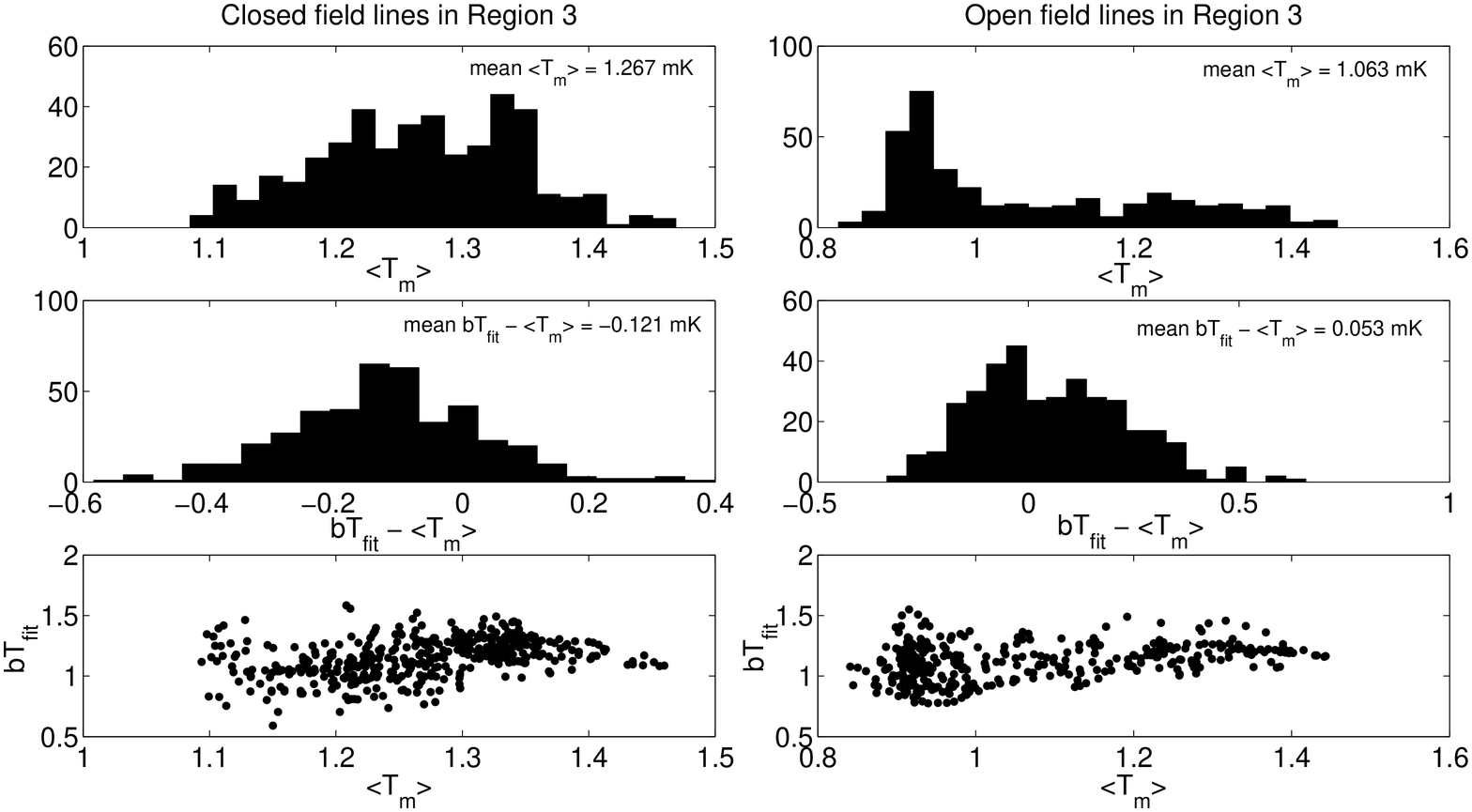}
\caption{Statistical analysis of the result of tracing individual closed (left
panels) and open (right panels) field lines $i$ in Region \R{3}. Top: histograms
of $\left<\Tm\right>^{(i)}$. Middle: histograms of $\bb\Tfit^{(i)}$. Bottom:
scatter plots of $\bb\Tfit^{(i)}$ \textit{versus} $\left<\Tm\right>^{(i)}$.}
\label{linetrace_R3}
\end{figure}

We performed the same analyses for regions \R{2} and \R{3}, and the results are
shown in Figures  \ref{linetrace_R2} and \ref{linetrace_R3}, respectively. For
the closed field lines, the top-left histograms show a mean fractional
difference [$\dT^{(i)}/\left<\Tm\right>^{(i)}$] of $-4$ and $-9\%$, for regions
\R{2} and \R{3} respectively. These numbers are quite similar to (and consistent
in sign with) those of $\dT/\left<\Tm\right>$ in Table \ref{tab1} for \R{2}-C
and \R{3}-C, which are $-1$ and $-6\%$ respectively. A negative value for
$\dT^{(i)}$ may be a direct indicator that $\Te>\Tions$, but it could also mean
that the loops are inherently dynamic on time scales not resolved by the
method.

{The top-right histogram of Figure  \ref{linetrace_R2} shows two distinct
populations of open field lines, centered around different mean temperatures,
larger and lower than $\approx$1 MK. A similar characteristic is seen in the
top-right histogram of Figure  \ref{linetrace_R3}. Consistently, in the top
panel of Figure \ref{LDEM_Tm_maps} the open part of region \R{2} shows $\Tm>1$
MK (yellow) and $<1$ MK (orange) voxels, and the same is seen in the open part
of region \R{3}. As regions \R{2}-O and \R{3}-O are almost exclusively populated
by $\Tm<1$ MK voxels, to compare with their results in Table 1 we consider now
only the population $\left<\Tm\right>^{(i)}<1$ MK of the corresponding
histogram. For that subset we find a mean fractional difference
$\dT^{(i)}/\left<\Tm\right>^{(i)}$ of order $+14\%$, both in region \R{2} and
\R{3}. This value compares quite well with the value
$\dT/\left<\Tm\right>=+12\%$ in Table \ref{tab1} for \R{2}-O, and is somewhat
smaller than the value $\dT/\left<\Tm\right>=+23\%$ value for \R{3}-O. In
summary, both the results of Table 1 and of the histograms for the magnetic
structures, reveal a consistent trend for the open subpolar regions to show
larger $\dT/\left<\Tm\right>$ values than the closed regions within the streamer
core.}

\subsection{Comparative Quantitative Analysis of CR-2068 and
CR-2077}\label{resultsB}

As qualitatively described in the previous section, the DEMT analysis of CR-2068
and CR-2077 (VFM10) reveals many similarities between both periods, as well as
differences. We include in this section a quantitative comparison of the two
periods, both belonging to the Solar Cycle 23 extended minimum phase. The WHI
CR-2068 (March\,--\,April 2008) period showed increased magnetic activity with
respect to CR-2077 (November\,--\,December 2008). This is due to the later being
closer to the absolute minimum of sunspot number for Solar Cycle 23, which
occurred in December 2008 according to the National Oceanic and Atmospheric
Administration Space Weather Prediction Center
(\url{http://www.swpc.noaa.gov/SolarCycle}). This is clearly manifested  by
the presence of the AR complex in the southern hemisphere near the equator in
CR-2068, that shows an overall more complex magnetic topology than CR-2077.

We begin by quantifying the relative diﬀerence in the two Carrington rotations
in a global way.  {Since the reconstructions from both periods are
subject to extremely similar systematic errors, the relative differences between
the two periods can be quantified much more precisely than absolute quantities.}
Inspection of the FBE Carrington maps for CR-2068 and CR-2077 (VFM10) shows that
both periods exhibit the most quiet characteristics throughout the CL range 0 to
160\deg. We use that range for this global analysis, as at higher CLs CR-2068
shows an AR complex. We also limit the analysis to the latitudinal range -75 to
+75\deg, to avoid the PSF contaminated regions of the CHs. In this coordinate
range, and with the aid of the PFSS model, we label the tomographic grid voxels
as open or closed, belonging to the streamer core and streamer-leg/subpolar
regions, respectively. 

For each given physical quantity of interest $Q$, we computed its average value
at a given height in the open and closed regions for both periods. We then
computed the ratio $R_Q\equiv \left<Q\right>_{\rm CR-2068}/\left<Q\right>_{\rm
CR-2077}$, for the open and closed regions separately. {At three different
heights of the tomographic grid (spherical shells of width $\Delta
r=0.01\mrSun$), Table \ref{tab2} shows, for each region, the ratios of: the
areas [$A$], the average FBEs [$\zeta_k$] in each of the three coronal bands,
the average LDEM electron density [$N_e$], the total mass in the shell [$M = \mu
m_H N_e A \Delta r$], the average electron mean temperature [$\Tm$], and the
average electron temperature spread [$\WT$].}

\begin{table}
\begin{tabular}{ccccccccc}
\hline
Height [\rsun]  & $R_{A}$ &
$R_{\zeta 171}$& $R_{\zeta 195}$& $R_{\zeta 284}$& 
$R_{N_e}$ & $R_{M}$& $R_{Tm}$&$R_{W_T}$\\
\hline
Closed Region\\
1.035 &   0.96 &   1.01 &   0.93 &   0.94 &   0.99 &   0.96 &   0.98 &
1.04\\
1.075 &   0.93 &   1.05 &   0.93 &   0.90 &   1.00 &   0.92 &   0.97 &  
1.00\\
1.135 &   0.90 &   1.05 &   0.98 &   0.89 &   1.01 &   0.90 &   0.99 &  
1.07\\
\hline
Open Region\\
1.035 &   1.24 &   1.09 &   0.98 &   0.75 &   1.07 &   1.32 &   1.07 &  
1.16\\
1.075 &   1.36 &   1.10 &   0.98 &   0.74 &   1.06 &   1.45 &   1.03 &  
1.05\\
1.135 &   1.38 &   1.08 &   0.94 &   0.75 &   1.04 &   1.44 &   1.01 &  
1.09\\
\hline
\end{tabular}
\caption{\footnotesize  Ratios $R_Q\equiv \left<Q\right>_{\rm
CR-2068}/\left<Q\right>_{\rm CR-2077}$ of different quantities (see text), for
the open and closed regions separately, at three heights.
\vspace{-1mm}}
\label{tab2}
\end{table}

In the three analyzed heights, the CR-2068 closed (streamer) area is 4 to 11\%
smaller than during CR-2077. {This is consistent with the slightly
larger latitudinal extent of the CR-2077 streamer core, which can also be seen
by comparing the location of the open/closed boundaries in Figure 1 with the
corresponding Figure 1 in VFM10. The average electron density in both periods is
about the same, and the lower total mass of the CR-2068 streamer is then due to
its smaller area. The average electron temperature in the CR-2068 streamer is
2\% lower, while the temperature spread is about 4\% larger.}

{In the three analyzed heights, the area of the CR-2068 open regions (which we
recall it includes only up to latitudes $\pm 75\mdeg$) is 24 to 38\% larger than
during CR-2077, while the average electron density is about $\approx$6\% larger.
The larger total mass of the CR-2068 open regions is then due to a contribution
of both factors, being the larger areas the dominant one.} The average electron
temperature in the CR-2068 open regions is about 4\% larger, while the
temperature spread is about 10\% larger.

Combining DEMT with the PFSS model, we computed at each voxel the plasma $\beta$
parameter for both periods. As shown in Figure \ref{beta_maps}, some localized
regions show very high (say $\beta > 5$) values within the closed region. As the
great majority (more than 80\%) of the voxels have $\beta < 5$, by taking the
median (instead of the mean) of the $\beta$ value in each region we obtain a
representative value not considering the highest values. For the considered
range of coordinates, Table \ref{tab3} shows the typical value of $\beta$ for
the closed (C) and open (O) regions. As already pointed out, the plasma $\beta$
is of order one within the streamer core, and much less than one in the
surrounding open regions, with both periods showing a similar scenario. The most
notable difference between both periods is found in the open regions, for which
CR-2077 shows lower values relative to CR-2068. We find then a slightly enhanced
closed-to-open $\beta$ contrast for CR-2077 respect to CR-2068.

\begin{table}
\begin{tabular}{ccccc}
\hline
$r/\mrSun$  &   C-2068 &  C-2077 & O-2068 & O-2077 \\
\hline
1.035&   0.99&   1.19&   0.14&   0.08\\
1.075&  1.43 &  1.71&    0.13&   0.05 \\
1.135&   1.78&   2.11&   0.25&   0.10\\
\hline
\end{tabular}
\caption{\footnotesize Typical plasma $\beta$ value in the closed (C) and open
(O) regions, for both periods, at three heights.\vspace{-1mm}}
\label{tab3}
\end{table}

The global analysis made so far compares average properties between the two
periods, for the closed and open voxels separately. The population of open
voxels sampled locations that are neighboring the open/closed boundary (streamer
legs) as well as higher subpolar latitudes. As there is an important latitudinal
density gradient across the streamer leg/subpolar region, the results of the
analysis of all the open voxels represent average properties of those regions.
To analyze the properties of the subpolar region only, we compare now the
results of the \R{i}-O regions in Table \ref{tab1} with the analysis we
performed in VFM10 for the CR-2077 subpolar region. In a similar way we will
compare the results for the streamer core region SC in Table \ref{tab1} with the
corresponding streamer core center results for CR-2077.

{To compare the dependence with height of the electron density in the streamer
core of both periods, we selected from CR-2077 reconstruction the same range of
latitude and longitude indicated as SC in Table I (which was also a quiet-Sun
part of the streamer region in that period),} and performed a similar
hydrostatic fit. We found almost the same average basal density $N_{e,0}^{\rm
(CR-2077)}=2.24\times 10^8\, {\rm cm^{-3}}$, and a scale height $\l^{\rm
(CR-2077)}=0.0829$ \rsun, a value 3\% smaller than the corresponding value for
CR-2068. 
The dependence with height of the electron density in the subpolar region is
also very similar to our result for CR-2077. In that case, averaging over the
subpolar region of both hemispheres, we found a basal density $N_{e,0}^{\rm
(CR-2077)}=1.04\times 10^8\, {\rm cm^{-3}}$, which is 4\% smaller than the
average of the basal densities of the three \R{i}-O regions. The average scale
height of the subpolar regions in CR-2077 $\l^{\rm (CR-2077)}=0.0828$ \rsun,
which is about equal to the average of the scale heights of the three
\R{i}-O regions.

In the quiet regions studied in Table \ref{tab1}, the dependence with height of
the mean electron temperature [$\Tm(r)$ shown in Figure \ref{tomo_height_Tm}]
exhibits variations of order 10\% over the analyzed height range, similarly to
what we found for CR-2077 (Figure 11 of VFM10). For both periods, the streamer
core shows similar temperatures, and in both cases the dependence of the
electron temperature with height [$\Tm(r)$] exhibits a global minimum, around
1.08
and 1.12 \rsun, for CR-2068 and CR-2077 respectively. In the subpolar regions,
both periods consistently showed mean electron temperatures in the range 0.85 to
1.0 MK. In subpolar regions \R{1}-O and \R{3}-O of CR-2068, and in the subpolar
regions of CR-2077, the temperature dependence with height consistently showed a
global minimum of about 0.85 MK at about 1.06 \rsun, in all cases, above where
we observe a monotonic increase of temperature with height, reaching about 1 MK
at 1.225 \rsun, in all cases.

Except for localized regions, such as the AR complex, the LDEM moments $\Tm$ and
$\WT$ are not related to each other in a simple way. The spatial distribution of
$\WT$ is more complicated in CR-2068 than what we found in CR-2077. For this
later period we found a simpler transition of larger to lower $\WT$ values,
generally located some degrees in latitude outside the magnetically open/closed
boundary.

\section{Discussion and Conclusions}\label{conclusions}

We studied the WHI CR-2068 corona by means of a dual-spacecraft DEMT and a PFSS
model, in a similar way to our previous analysis of the period CR-2077 (VFM10).
Both periods belong to the Solar Cycle 23 extended minimum phase. The
tomographic reconstruction allowed the LDEM analysis of the corona in the height
range 1.03 and 1.23 \rsun. Taking moments of the obtained LDEM in each
tomographic cell, we produced 3D maps of the distribution of the electron
density [$\Ne$], mean electron temperature [$\Tm$], and electron temperature
spread
[$\WT$]. For interpretation of the DEMT results in terms of the corona magnetic
structure, we also used a PFSS model of the solar corona based on MDI/SOHO
magnetograms of the same period. 

The comparison of the CR-2068 and CR-2077 streamer structures indicates that the
volume of the closed region increased as the global minimum of activity was
reached, being the streamer area 4 to 10 \% larger between heights 1.035 and
1.135 \rsun, and exhibiting in both periods similar densities and temperatures.
On the other hand, the streamer legs and subpolar open regions have a
considerably larger volume during CR-2068, and are populated by a slightly
hotter and more spread temperature distribution, respect to CR-2077. The
$R_{N_e}$ values show that the closed-to-open density contrast is in average
$\approx$6\% larger for CR-2077. The larger streamer volume of CR-2077 can be
understood then as the result of its expansion due to a larger gas pressure in
the streamer core relative to its open surroundings, which also suggests a
larger cusp height. Consistently, Table 2 shows a closed-to-open $\beta$
contrast for CR-2077 that is of order twice that of CR-2068.

Our results are consistent then with a streamer region characterized by a
relatively large gas pressure, surrounded by open regions of low $\beta$ acting
as a magnetic container upon the streamer core. Similar results were found by Li
\etal{} (1998), who studied a previous solar minimum streamer (July 1996) with
data provided by the \textit{Ultraviolet Coronagraph Spectrometer} (UVCS/SOHO,
Kohl \etal, 1995) and the \textit{Soft X-Ray telescope} (SXT) of the
\textit{Yohkoh} mission. Using also a potential field magnetic model, they
estimated $\beta$ within the streamer core, and found values $\beta\approx 5$ at
1.15 \rsun\ and $\approx 3$ at 1.5\rsun. High values of $\beta$ in streamers
were also found in MHD models that include heat and momentum deposition in the
corona (Wang \etal, 1998; Suess \etal, 1996), or empirically prescribed
temperatures (V\'asquez \etal, 2003). In our 3D maps the boundaries of the
$\beta>1$ regions within the streamer core are large scale polarity inversion
lines. This is particularly clear in the 1.075 and 1.135 \rsun\ Carrington maps
of Figure \ref{beta_maps}.

The hydrostatic fit to the electron density variation with height in the
streamer core of the WHI period is found to have a basal density of about
$2.23\times 10^8\, {\rm cm^{-3}}$ and a scale height of $0.086\,\mrSun$, with
similar values for the CR-2077 period. These results can be compared with
studies of the previous solar minimum. In the Gibson \etal{} (1999) analysis of
the Whole Sun Month (WSM, August 1996) coronal streamer, the authors consider a
similar range of latitudes, [-27\deg,+27\deg], and derive electron densities in
the height range 1.0 to 1.2 \rsun\ using EUV line ratios computed from the
\textit{Coronal Diagnostic Spectrometer} (CDS) onboard SOHO. Gibson's densities
are larger than ours by a roughly constant factor of 2.05. {Differences may be
partly due to physical changes between both minima and/or to our assumed Fe
abundance being too large, and also due to the CDS diagnostic uncertainty. Also,
their study used outdated CHIANTI data and did not include radiative excitation.
In addition, their diagnostic is on a part of the curve where the line ratio is
not very density sensitive, so the observational uncertainties can easily lead
to a factor of two or three. Thus, their results are consistent with ours, to
within the large uncertainties.}

Li \etal{} (1998) studied the July 1996 equatorial streamer, using
\textit{Yohkoh}/SXT data at the height range of our study, and using SOHO/UVCS
data at larger heights. They found electron densities at 1.15 \rsun\ to be about
2.5 times larger  than ours, and at 1.5 \rsun\ to be about 2.7 times larger than
the extrapolation of our hydrostatic fit for the streamer-core region.  It
should be noted that both results are subject to quite important sources of
uncertainty, such as the very broad response functions in the case of SXT
observations, inaccuracies in the Ly-$\beta$ disk intensity, and the
collisional/radiative ratio of the Ly-$\beta$ line from which the density is
determined with UVCS observations. 

{Our results for density are quite consistent with studies based on data taken
by the SOHO Ultraviolet Measurement of Emitted Radiation (SUMER). For example,
using SUMER data taken during a special ``roll'' maneuver,} Feldman \etal{}
(1999) analyzed the November 1996 equatorial streamer along its axis structure.
Their estimated electron density values at 1.04 and 1.2 \rsun, are respectively
27\% and 20\% larger than those predicted by our hydrostatic fit of region SC in
Table I, differences that are well within the SUMER diagnostic uncertainties. It
should also be noted that their results, as well as those from SXT and UVCS
studies mentioned above, are affected by LOS integration. {Wilhelm \etal{}
(2002) analyzed equatorial streamers observed by SUMER in 1998, during the
rising phase of Solar Cycle 23. At the base of the streamers, they derived an
electron density of $2\times10^8{\rm cm}^{-3}$ at 1.01 \rsun, which agrees very
well with our results).}

Using hydrostatic fitting technique, Gibson \etal{} (1999) obtained an average
electron temperature of 1.25 MK below 1.2 \rsun, that is similar to the
$b\Tfit=1.16$ MK value we obtain for the streamer core in Table \ref{tab1}.
Feldman \etal{} (1999) also found the streamer core axis to be quite isothermal,
with a similar typical temperature of about 1.3 MK. Li's estimated electron
temperatures at that height in the streamer core are of order 1.7 MK, which are
considerably larger than the typical values we find, but have additional large
uncertainties due to the XRT response. {We found an average $\Tm$ value of about
1.13 MK in the (closed) streamer core, with the largest values located near the
open/closed boundary, and about 0.93 MK in the (open) subpolar latitudes (Figure
\ref{tomo_height_Tm}). Results from the analysis of solar minimum coronal
streamers based on UVCS data are generally consistent with the closed region
being in collisional thermal equilibrium (Strachan \etal, 2002). Analyses of
streamer core regions, based on UVCS spectra Ly-$\alpha$ and O\,{\sc vi} line
widths, have indicated kinetic temperatures similar to $T_e$ values derived from
ionization balance models (Raymond \etal, 1997; Raymond, 1999), as well as to
the UVCS measurement of the Thomson-scattered Ly-$\alpha$ width (Fineschi \etal,
1998), a direct measure of the electron temperature. At 1.5 \rsun, different
studies based on UVCS data of solar minimum streamers derived values of $T_e$ in
the range 1.1 to 1.5 MK (Kohl \etal, 2006; Uzzo \etal, 2007, and references
therein), which are roughly consistent with our results.}

{The Carrington maps of the LDEM mean electron temperature (Figure
\ref{LDEM_Tm_maps}, as well as the corresponding maps for CR-2077 in VFM10) show
in the streamer core typical values of order 1 MK, while the closed regions at
mid latitudes near the open/closed boundary show enhanced values of up to about
1.5 MK.  In their 1998 streamer study based on SUMER data, Wilhelm \etal{}
(2002) found a very similar scenario, with electron temperatures close to 1 MK
at low latitudes, and of 1.4 MK in closed regions at mid latitudes (see also
Moses \etal, 1994; Guhathakurta and Fisher, 1994).}

For several quiet regions, both in the streamer core and the subpolar latitudes,
we performed a least-squares fit of a hydrostatic solution to the LDEM electron
density dependence with height. In the subpolar regions, the hydrostatic fits of
CR-2068 and CR-2077 showed very similar trends, with an average basal density
$N_{e0} = 1.06 \times 10^8 {\rm cm^{-3}}$ and scale height $\l=0.083\mrSun$.
This average fit gives electron density values which are 66 and 52\% lower than
CH inter-plume values measured at heliocentric heights 1.03 and 1.26 \rsun,
respectively, derived from spectroscopic data taken by SUMER (Banerjee \etal,
1998) during the previous solar minimum.  Again, the difference may be partly
due to our assumed Fe abundance being too high, and also due to the fact that
both open regions correspond to different latitudinal ranges. Over the same
height range, our analysis of the subpolar regions in both periods shows
electron temperatures in the range 0.85 to 1.0 MK. {Wilhelm \etal{} (1998)
determined coronal hole electron temperatures in the range 0.75 to 0.88 MK from
SUMER observations. A recent study of stereoscopically reconstructed polar
plumes indicated electron temperatures in the range 0.85 to 0.90 MK, derived
from SUMER data (Feng \etal, 2009).  The fact that the open regions we studied
are at subpolar latitudes may be the cause of our temperature values being up tp
10\% larger than in these CH studies.}

Within each selected region, we analyzed the average dependence with height of
the LDEM $\Ne$, and compared the fit scale height temperature $\Tfit$ with the
corresponding LDEM $\Tm$, which we take to be a measure of the true electron
temperature. We have also traced individual magnetic-field lines through the
tomographic computational grid, and applied an hydrostatic fit analysis to the
LDEM density along every individual field line. We performed a statistical
analysis of the results along the individual field lines, separately for the
closed and the open field lines. Our analyses show that the density dependence
with height along individual magnetic-field lines in each region is in general
well represented by the results summarized in Table \ref{tab1}. We draw the
following conclusions:

\begin{enumerate}

\item 
In the magnetically closed regions of the streamer core, our results are
consistent with a plasma regime of hydrostatic isothermal equilibrium, allowing
for either over-density or $\Te > \Tions$, with a temperature difference of up
to about 10\%. This is seen both in the streamer core central latitudes, as well
as closer to the boundary with the surrounding open magnetic structures.

\item On the open field lines, we found considerably larger values of
$b\Tfit-\left<\Tm\right>$, the difference between the scale-height temperature
and the LDEM electron temperature, than those in the streamer core.
Nevertheless, the hydrostatic fits have small residuals, indicating one or both
of these scenarios:

\begin{enumerate}

\item Pressure mechanisms other than thermal are acting, and these mechanisms
are approximately linear in the density.

\item Isothermality among species is not met. In this case, our analysis
indicates that $T_{\rm ions}>\Te$, and that the temperature ratio can
typically reach values up to order $T_{\rm ions}/\Te\approx1.5$ at subpolar
latitudes.

\end{enumerate}

\end{enumerate}

Ion excess temperature ($\Tions>\Te$) in quiet- and active-Sun regions have been
found observationally by several authors (Seely \etal, 1997; Tu \etal, 1998).
Landi (2007) analyzed quiet region off-disk spectra observed by SUMER, for three
different periods of the last solar cycle (1996, 1999, 2000).  Landi estimated
the possible temperature ranges of several ions, between the limb and 1.25
\rsun. In all cases, the author estimated electron temperatures in the range
1.25 to 1.35 MK, and a systematic trend for $\Tions>\Te$, with indication of a
larger average ions excess temperature for increasing activity of the Sun.
Typical temperature ratios [$\Tions/\Te$] were found in the range one to two,
with
peak values of up to order three, which is consistent with our findings. In
their
stereoscopically reconstructed CH plumes study, Feng \etal{} (2009) reported
density scale height temperatures to be about 70\% larger than their electron
temperature values (derived from SUMER observations), assigning the discrepancy
to ion excess temperatures.

To further explore our conclusions, we plan to develop an extensive analysis of
other solar regions, refining the comparisons by developing non-potential field
extrapolation for selected regions. Another important improvement (currently
under development) is the incorporation of the PSF deconvolution of the images
used for tomography, allowing us to extend DEMT analysis to coronal holes. The
global character of DEMT analysis make it suitable to serve as observational
constraint to global coronal models. As an example, we have recently used the
CR-2077 DEMT results to provide electron density and temperature basal boundary
conditions for a two-temperature MHD model of the solar wind (van der Holst,
2010). The usefulness of the DEMT results in general, and as constraint for
models in particular, will improve much by the inclusion of the PSF
deconvolution.

Immediate future work includes the application of the DEMT technique to use the
six Fe bands of the SDO/AIA instrument.  Taking advantage of the increased
number of bands and temperature coverage provided by SDO/AIA, we aim to refine
the LDEM determination. Finally, we also plan to implement time-dependent DEMT
through the application of the Kalman-filtering method (Frazin \etal, 2005;
Butala \etal, 2008).

\section*{Acknowledgments}

We are grateful to the referee for the careful reading of our manuscript and the
enriching suggestions, which considerably improved the exploitation of results
and the organization of their exposition. We thank Enrico Landi for useful
discussions on the comparison of DEMT results with other density and temperature
diagnostics. We thank Jean-Pierre Wuelser for his priceless assistance with the
EUVI data. We thank Huw Morgan for his work with the IfA solar data catalogs.
This research was supported by NASA Heliophysics Guest Investigator award
NNX08\-A\-J\-0\-9G to the University of Michigan. A.M.V. also acknowledges the
ANPCyT PICT/33370-234 grant to IAFE for partial support. W.B.M.IV acknowledges
NASA Grant Number LWS NNX09AJ78G.

The STEREO/SECCHI data used here are produced by an international consortium of
the Naval Research Laboratory (USA), Lockheed Martin Solar and Astrophysics Lab
(USA), NASA Goddard Space Flight Center (USA) Rutherford Appleton Laboratory
(UK), University of Birmingham (UK), Max-Planck-Institut f\"ur
Sonnensystemforschung (Germany), Centre Spatiale de Li\`ege (Belgium), Institut
d'Optique Théorique et Applique\'e (France), Institut d'Astrophysique Spatiale
(France).

\clearpage

\end{article}
\end{document}